\documentstyle[preprint,tighten,aps,epsfig]{revtex}
\begin{document}
\preprint{MKPH-T-99-1}
\input{psfig}
\draft
\tighten
\title{Real and Virtual Compton Scattering 
at Low Energies\thanks{Lectures at the 11th Indian-Summer School on 
Intermediate Energy Physics {\em Mesons and Light Nuclei},
Prague, September 7 - 11, 1998, Czech Republic.}}
\author{S.\ Scherer}
\address{Institut f\"ur Kernphysik,
Johannes Gutenberg-Universit\"at,
D-55099 Mainz, Germany}
\date{11.1.1999}
\maketitle
\begin{abstract}
   These lectures give a pedagogical introduction to real and virtual
Compton scattering at low energies. 
   We will first discuss real Compton scattering off a point particle
as well as a composite system in the framework of nonrelativistic
quantum mechanics.
   The concept of electromagnetic polarizabilities is introduced.
   We then address a description of the Compton-scattering tensor
within quantum field theory with particular emphasis on the
derivation of low-energy theorems.
   The importance of a consistent treatment of hadron structure
in the use of electromagnetic vertices is stressed.
   Finally, the reader is introduced to the notion of generalized 
polarizabilities in the rapidly expanding field of 
virtual Compton scattering. 
\end{abstract}

\section{Introduction}
   The discovery of the Compton effect \cite{Compton_23,Debye_23}, 
{\em i.e.}, the scattering of photons off electrons, and its explanation 
in terms of conservation of energy and momentum in the collision
between a single light quantum with an electron is regarded as one of the key 
developments of modern physics \cite{Stuewer_77}.
   In atomic physics, condensed matter physics, and chemistry Compton 
scattering is nowadays an important tool of investigating the momentum
distribution of the scattering electrons in the probe.
   The inclusion of the electron spin in the calculation of the 
Compton-scattering cross section by Klein and Nishina \cite{Klein_29} has 
become one of the textbook examples of applying quantum electrodynamics
at lowest order.

   In the realm of strong-interaction physics, the potential of using
Compton scattering as a method of studying properties of particles was
realized in the early fifties.
   The influence of the {\em anomalous} magnetic moment of the proton on 
the Compton-scattering cross section was first discussed by
Powell \cite{Powell_49}.
   The derivation of low-energy theorems (LETs), {\em i.e.}, model-independent
predictions based upon a few general principles, became an important 
starting point in understanding hadron structure \cite{Thirring_50,%
Kroll_54,Low_54,GellMann_54}.
   Typically, the leading terms of the low-energy amplitude for a given 
reaction are predicted in terms of global, model-independent properties 
of the particles.
   LETs provide an important constraint for models or theories of hadron 
structure: unless these general principles are violated, the 
predictions of a low-energy theorem must be reproduced.
   Furthermore, LETs also provide useful constraints
for experiments as they define a reference point for the precision which 
has to be achieved in experimental studies designed 
to distinguish between different models.

     Based on the requirement of gauge invariance, Lorentz invariance, 
crossing symmetry, and the discrete symmetries, the low-energy theorem for 
Compton scattering (CS) of real photons off a nucleon 
\cite{Low_54,GellMann_54} uniquely specifies the terms in the 
low-energy scattering amplitude up to and including terms linear in the 
photon momentum.
   The coefficients of this expansion are expressed in terms of global
properties of the nucleon: its mass, charge, and magnetic moment.
   Terms of second order in the frequency, which are not determined by
this theorem, can be parameterized in terms of two new structure
constants, the electric and magnetic polarizabilities of the
nucleon.
   These polarizabilities have been the subject of numerous experimental
and theoretical studies as they determine the first information on 
the compositeness or structure of the nucleon specific to Compton scattering.

   As in all studies with electromagnetic probes, the possibilities to
investigate the structure of the target are much greater if virtual photons
are used, since the energy and three-momentum of the virtual photon
can be varied independently.
   Moreover, the longitudinal component of current operators entering 
the amplitude can be studied.
   The amplitude for virtual Compton scattering (VCS) off the proton
is accessible in the reactions $e^-p\to e^-p\gamma$ and 
$\gamma p\to p e^- e^+$.
   In particular, the first process has recently received considerable 
interest as it allows to investigate generalizations of the RCS
polarizabilities to the spacelike region, namely, the
so-called generalized polarizabilities \cite{Guichon_95}.
   
   The purpose of these lectures is to provide an {\em introduction} to the
topics of real and virtual Compton scattering.
   The material is organized in three chapters.
   We start at an elementary level in the framework of nonrelativistic
quantum mechanics and discuss basic features of Compton scattering.
   Then, a covariant treatment within quantum field theory
is discussed with particular emphasis on a consistent treatment of
compositeness in the use of electromagnetic vertices.
   In the last chapter, the reader is introduced to the rapidly
expanding field of virtual Compton scattering. 
   
   In preparing these lectures, we have made use of the excellent 
pedagogical reviews on hadron polarizabilities of 
Refs.\ \cite{Friar_89,Holstein_90,Holstein_92}.
   A vast amount of more detailed information is contained in
Refs.\ \cite{Petrunkin_81,Lvov_93}.
   An overview of the current status of experimental and theoretical 
activities on hadron polarizabilities can be found in Ref.\ \cite{WGS_98}.
   Finally, for a first review on virtual Compton scattering the
reader is referred to Ref.\ \cite{Guichon_98}.

\section{Compton scattering in nonrelativistic quantum mechanics}
\subsection{Kinematics and notations}
\begin{figure}
\caption[test]{\label{figurekin} 
Compton scattering: kinematics.}
\begin{center}
\epsfig{file=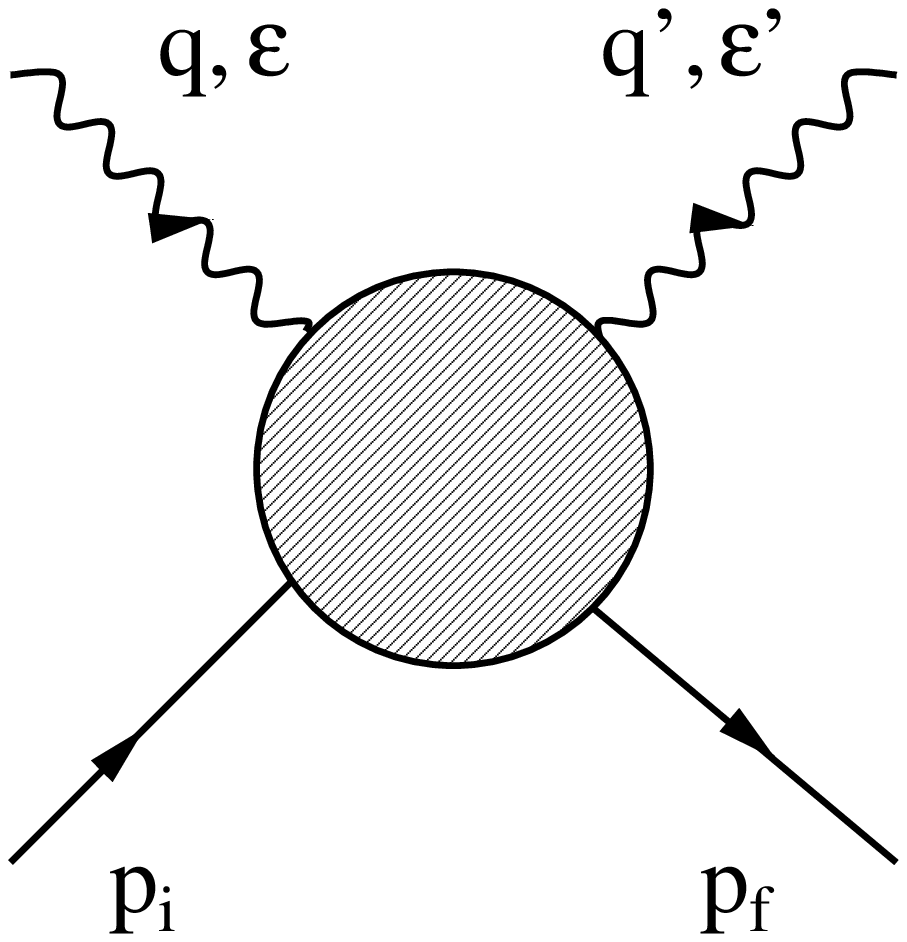,width=6cm}
\end{center}
\end{figure}

   We will first discuss real Compton scattering (RCS), for which
$q^2=q'^2=q\cdot\epsilon=q'\cdot\epsilon'=0$.
   The kinematical variables and polarization vectors are defined in
Fig.\ \ref{figurekin}.
   As a result of translational invariance in space-time, the total 
three-momentum and energy, respectively, are conserved,
\begin{equation}
\label{epcons}
\vec{p}_i+\vec{q}=\vec{p}_f+\vec{q}\,',\quad E_i+\omega=E_f+\omega', 
\end{equation}
   where the energy of the particle is given by
$E=\frac{\vec{p}\,^2}{2M}$ or $E=\sqrt{M^2+\vec{p}\,^2}$ depending on
whether one uses a nonrelativistic or relativistic framework.
   For the description of the RCS amplitude one requires two kinematical 
variables, {\em e.g.}, the energy of the initial photon, 
$\omega$, and the scattering angle between the initial photon and the 
scattered photon, $\cos(\Theta)=\hat{q}\cdot\hat{q}\,'$.
   The energy of the scattered photon in the lab frame
is given by
\begin{equation}
\label{omegap}
\omega'=\frac{\omega}{1+\frac{\omega}{M}[1-\cos(\Theta)]},
\end{equation}
   if use of relativistic kinematics is made.
   From Eq.\ (\ref{omegap}) one obtains the well-known result for
the wavelength shift of the Compton effect, 
$\Delta \lambda=(4\pi/M) \sin^2(\Theta/2)$.

\subsection{Nonrelativistic Compton scattering off a point particle}
   In order to set the stage, we will first discuss, in quite some detail,
Compton scattering of real photons off a free point particle of mass $M$ 
and charge $e>0$ within the framework of nonrelativistic quantum mechanics. 
   First of all, this will allow us to introduce basic concepts such as gauge 
invariance, photon-crossing symmetry as well as discrete symmetries.
   Secondly, the result will define a reference point beyond which the 
structure of a composite object can be studied.
   Finally, this will also allow us to discuss later on, where a relativistic 
description departs from a nonrelativistic treatment.

   Consider the Hamiltonian of a single, free point particle of mass $M$
and charge $e>0$,\footnote{Except for a few cases, we will use the
same symbols for quantum-mechanical operators such as $\hat{\vec{p}}$
and corresponding eigenvalues $\vec{p}$.}
\begin{equation}
\label{h0}
H_0=\frac{\vec{p}\,^2}{2M}.
\end{equation}
   The coupling to the electromagnetic field, $A^\mu(\vec{x},t)=
(\Phi(\vec{x},t),\vec{A}(\vec{x},t))$, is generated by the well-known
minimal-substitution procedure\footnote{We use Heaviside-Lorentz units,
$e>0$, $\alpha=e^2/4\pi\approx1/137$.}
\begin{equation}
\label{minsub}
i\frac{\partial}{\partial t}\mapsto i\frac{\partial}{\partial t} - e 
\Phi(\vec{x},t),\quad
\vec{p}\mapsto\vec{p}-e \vec{A}(\vec{x},t),
\end{equation}
resulting in the Schr\"{o}dinger equation
\begin{equation}
\label{schr}
i\frac{\partial \Psi(\vec{x},t)}{\partial t}=
[H_0+H_I(t)]\Psi(\vec{x},t)=
[H_0+H_1(t)+H_2(t)]\Psi(\vec{x},t)=
H(\Phi,\vec{A})\Psi(\vec{x},t),
\end{equation}
where
\begin{equation}
H_1(t)=-e\frac{\vec{p}\cdot\vec{A}+\vec{A}\cdot{\vec{p}}}{2M}+e\Phi,
\quad
H_2(t)=\frac{e^2}{2M}\vec{A}\,^2.
\end{equation}
   Gauge invariance of Eq.\ (\ref{schr})  means that
\begin{equation}
\Psi'(\vec{x},t)=\exp[-ie\chi(\vec{x},t)]\Psi(\vec{x},t)
\end{equation}
is a solution of
\begin{equation}
i\frac{\partial \Psi'(\vec{x},t)}{\partial t}=
H(\Phi+\dot{\chi},\vec{A}-\vec{\nabla}\chi)\Psi'(\vec{x},t),
\end{equation}
provided $\Psi(\vec{x},t)$ is a solution of Eq.\ (\ref{schr}).
   In other words, Eq.\ (\ref{schr}) remains invariant under a gauge 
transformation
\begin{equation}
\Psi\mapsto\exp[-ie\chi(\vec{x},t)]\Psi, \quad A^\mu\mapsto A^\mu+
\partial^\mu\chi.
\end{equation}
   For a discussion of gauge invariance in the context of nonrelativistic
reductions the interested reader is referred to Ref.\ \cite{Scherer_94}.    

   After introducing the interaction representation,
\begin{equation}
\label{intrep}
H_I^{int}(t)=e^{i H_0 t} H_I(t)e^{-i H_0 t},
\end{equation}
   the $S$-matrix element is obtained by evaluating the Dyson series
\begin{equation}
\label{dyson}
S = 1+ \sum_{k=1}^\infty \frac{(-i)^k}{k!}\int_{-\infty}^\infty
dt_1 \cdots dt_k \hat{T}\left[H^{int}_I(t_1)\cdots H^{int}_I(t_k)\right]
\end{equation}
   between $|i\!>\equiv|\vec{p}_i; \gamma(q,\epsilon)\!>$ and
$<\!f|\equiv \mbox{$<\!\vec{p}_f; \gamma(q', \epsilon')|$}$.
   In Eq.\ (\ref{dyson}), $\hat{T}$ refers to the time-ordering operator,
\begin{equation}
\hat{T}\left[A(t_1) B(t_2)\right]=A(t_1) B(t_2)\Theta(t_1-t_2)
+B(t_2) A(t_1)\Theta(t_2-t_1),
\end{equation}
   with a straightforward generalization to an arbitrary number of operators.
   We use second-quantized photon fields 
\begin{equation}
\label{sqpf}
<\! 0|A^\mu(\vec{x},t)|\gamma[q,\epsilon(\lambda)]\!>=
\epsilon^\mu(q,\lambda)N(\omega)e^{-iq\cdot x},
\end{equation}
   where $N(\omega)=[(2\pi)^3 2\omega]^{-1/2}$, and normalize the states as
\begin{equation}
\label{normstat}
<\!\vec{x}|\vec{p}\!>=\frac{e^{i\vec{p}\cdot\vec{x}}}{\sqrt{(2\pi)^3}}.
\end{equation}
   The part relevant to Compton scattering [${\cal O}(e^2)$] reads
\begin{equation}
\label{scomp}
S=-i\int_{-\infty}^\infty dt H^{int}_2(t)
-\int_{-\infty}^\infty dt_1 dt_2
H_1^{int}(t_1) H_1^{int}(t_2)\Theta(t_1-t_2),
\end{equation}
   where the first term generates the contact-interaction contribution 
or so-called seagull term:
\begin{eqnarray}
\label{scont}
S_{fi}^{cont}&=&
-i\frac{e^2}{2M}\int_{-\infty}^\infty dt <\!f|
e^{iH_0 t}\vec{A}\,^2(\hat{\vec{r}},t) e^{-iH_0 t}
|i\!>\nonumber\\\
&=&-i(2\pi)^4\delta^4(p_f+q'-p_i-q)\frac{1}{\sqrt{4\omega\omega'}(2\pi)^6}
\frac{e^2 \vec{\epsilon}\,'^\ast\cdot\vec{\epsilon}}{M}.
\end{eqnarray}
   In order to obtain Eq.\ (\ref{scont}), one first contracts the photon
field operators with the photons in the initial and final states, 
respectively,\footnote{Note the factor of 2 for two contractions.} 
then evaluates the time integral, and, finally, makes use of 
\begin{displaymath}
<\!\vec{p}\,'|f(\hat{\vec{r}})|\vec{p}\!>=\frac{1}{(2\pi)^3}\int
d^3r e^{i(\vec{p}-\vec{p}\,')\cdot\vec{r}}f(\vec{r})
\end{displaymath}
   with $f(\vec{r})=\exp[i(\vec{q}-\vec{q}\,')\cdot\vec{r}]$, to obtain
the three-momentum conservation.
   The second contribution of Eq.\ (\ref{scomp}) is evaluated by inserting a 
complete set of states
   between $H_1^{int}(t_1)$ and $H_1^{int}(t_2)$:
\begin{equation}
-\int_{-\infty}^\infty dt_1 dt_2 \Theta(t_1-t_2)
\int d^3p
<\! f|H^{int}_1(t_1)|\vec{p}\!><\!\vec{p}| H^{int}_1(t_2)|i\!>.
\end{equation}
   There are two distinct possibilities to contract the photon fields,
namely, $A_\nu(t_2)$ with $|\gamma(q,\epsilon)\!>$ and 
 $A_\mu(t_1)$ with $<\!\gamma(q',\epsilon')|$ and vice versa,
giving rise to the so-called direct and crossed channels, respectively.
   Evaluating the time dependence and making use of 
\begin{displaymath}
\int_{-\infty}^\infty dt_1 dt_2 \Theta(t_1 - t_2) 
e^{iat_1} e^{ibt_2}
=\frac{2\pi i\delta(a+b)}{a+i0^+}
\end{displaymath} 
    one obtains 
\begin{eqnarray}
\label{sppdccc1}
S_{fi}^{dc+cc}&=&-2\pi i\delta(E_f+\omega'-E_i-\omega)\int d^3p
\left(\frac{<\vec{p}_f|H^{em}_1|\vec{p}><\vec{p}|H^{abs}_1|\vec{p}_i>}{
E_f+\omega'-E(\vec{p})+i0^+}\right.\nonumber\\
&&\quad\left.
+\frac{<\vec{p}_f|H^{abs}_1|\vec{p}><\vec{p}|H^{em}_1|\vec{p}_i>}{
E_f-\omega-E(\vec{p})+i0^+}\right),
\end{eqnarray}
   where the superscripts $abs$ and $em$ refer to absorption and emission
of photons, respectively, and where the matrix elements are given by
\begin{eqnarray}
<\!\vec{p}|H^{abs}_1|\vec{p}_i\!>&=&-eN(\omega)
\delta^3(\vec{p}-\vec{q}-\vec{p}_i)
\left[\frac{(\vec{p}+\vec{p}_i)\cdot\vec{\epsilon}}{2M}-
\epsilon_0\right],\\
<\!\vec{p}_f|H^{em}_1|\vec{p}\!>&=&-eN(\omega')
\delta^3(\vec{p}_f+\vec{q}\,'-\vec{p})
\left[\frac{(\vec{p}_f+\vec{p})\cdot\vec{\epsilon}\,'^\ast}{2M}-
\epsilon_0'^\ast\right].
\end{eqnarray}
   Using the following convention 
\begin{displaymath}
{\cal T}_{fi}=\frac{1}{\sqrt{4\omega\omega'}(2\pi)^6}t_{fi},
\end{displaymath}
with $S=I+iT$ at the operator level and 
$<\! f|T|i\!>=(2\pi)^4\delta^4(P_f-P_i){\cal T}_{fi}$,
the final result for the $T$-matrix element reads
\begin{eqnarray}
t_{fi}&=&e^2\left\{-\frac{\vec{\epsilon}\,'^\ast\cdot\vec{\epsilon}}{M}
-\left[\frac{(2\vec{p}_f+\vec{q}\,')\cdot
\vec{\epsilon}\,'^\ast}{2M} -\epsilon_0'^\ast\right]
\frac{1}{E_f+\omega'-E(\vec{p}_f+\vec{q}\,')}
\left[\frac{(2\vec{p}_i+\vec{q})\cdot
\vec{\epsilon}}{2M}-\epsilon_0\right]\right.\nonumber\\
&&\left.\hspace{4em}
-\left[\frac{(2\vec{p}_f-\vec{q})\cdot\vec{\epsilon}}{2M} -\epsilon_0\right]
\frac{1}{E_f-\omega-E(\vec{p}_f-\vec{q})}
\left[\frac{(2\vec{p}_i-\vec{q}\,')\cdot
\vec{\epsilon}\,'^\ast}{2M}-\epsilon_0'^\ast\right]\right\}.
\end{eqnarray}
   Let us discuss a few properties of $t_{fi}$. 
\begin{itemize}
\item Gauge invariance: As a result of the gauge-invariance property
of the equation of motion, the result for true observables should
not depend on the gauge chosen.  
   In the present context, this means that the transition matrix element
is invariant under the replacement $\epsilon^\mu\to \epsilon^\mu 
+\zeta q^\mu$ (analogously for $\epsilon'$):
\begin{eqnarray*}
t_{fi}&\stackrel{\epsilon^\mu\to q^\mu}{\mapsto}&
e^2\left\{-\frac{\vec{\epsilon}\,'^\ast\cdot\vec{q}}{M}
-\left[\frac{(2\vec{p}_f+\vec{q}\,')\cdot
\vec{\epsilon}\,'^\ast}{2M} -\epsilon_0'^\ast\right]
\frac{1}{E_f+\omega'-E(\vec{p}_f+\vec{q}\,')}
\left[\frac{(2\vec{p}_i+\vec{q})\cdot\vec{q}}{2M}-\omega\right]\right.\\
&&\left.-\left[\frac{(2\vec{p}_f-\vec{q})\cdot
\vec{q}}{2M} -\omega\right]
\frac{1}{E_f-\omega-E(\vec{p}_f-\vec{q})}
\left[\frac{(2\vec{p}_i-\vec{q}\,')\cdot
\vec{\epsilon}\,'^\ast}{2M}-\epsilon_0'^\ast\right]\right\}\\
&\stackrel{(\ast)}{=}&
e^2\left\{-\frac{\vec{\epsilon}\,'^\ast\cdot\vec{q}}{M}
+\left[\frac{(2\vec{p}_f+\vec{q}\,')\cdot
\vec{\epsilon}\,'^\ast}{2M} -\epsilon_0'^\ast\right]
-\left[\frac{(2\vec{p}_i-\vec{q}\,')\cdot
\vec{\epsilon}\,'^\ast}{2M}-\epsilon_0'^\ast\right]\right\}\\
&=&0\quad\mbox{since}\quad2\vec{p}_f+\vec{q}\,'-2\vec{p}_i+\vec{q}\,'
=2\vec{q},
\end{eqnarray*}
   where, using energy conservation, in $(\ast)$ we inserted 
\begin{eqnarray*}
E_f+\omega'-E(\vec{p}_f+\vec{q}\,')&=&
-\left[\frac{(2\vec{p}_i+\vec{q})\cdot\vec{q}}{2M}-\omega\right],\\
E_f-\omega-E(\vec{p}_f-\vec{q})&=&
\left[\frac{(2\vec{p}_f-\vec{q})\cdot\vec{q}}{2M}-\omega\right].
\end{eqnarray*}

\item Photon-crossing symmetry: $t_{fi}$ is invariant under the 
simultaneous replacements
$\epsilon^\mu\leftrightarrow \epsilon'^{\mu\ast}$ and 
$q^\mu\leftrightarrow -q'^\mu$, {\em i.e.},
\begin{eqnarray*}
t_{fi}&\mapsto& e^2\left\{-\frac{\vec{\epsilon}\cdot\vec{\epsilon}\,'^\ast}{M}
-\left[\frac{(2\vec{p}_f-\vec{q})\cdot
\vec{\epsilon}}{2M} -\epsilon_0\right]
\frac{1}{E_f-\omega-E(\vec{p}_f-\vec{q})}
\left[\frac{(2\vec{p}_i-\vec{q}\,')\cdot
\vec{\epsilon}\,'^\ast}{2M}-\epsilon_0'^\ast\right]\right.\\
&&\left.\hspace{4em}-\left[\frac{(2\vec{p}_f+\vec{q}\,')\cdot
\vec{\epsilon}\,'^\ast}{2M} -\epsilon_0'^\ast\right]
\frac{1}{E_f+\omega'-E(\vec{p}_f+\vec{q}\,')}
\left[\frac{(2\vec{p}_i+\vec{q})\cdot
\vec{\epsilon}}{2M}-\epsilon_0\right]\right\}\\
&=&t_{fi}.
\end{eqnarray*}
\item Invariance under $e\mapsto -e$, {\em i.e.}, the Compton-scattering
amplitudes for particles of charges $e$ and $-e$ are identical.  
\item Under parity, $t_{fi}$ behaves as a scalar, {\em i.e.}, there
are no terms of, {\em e.g.}, the type
$\epsilon_{ijk} \epsilon_i \epsilon_j'^\ast q_k$.
\item Particle crossing, $(E_i,\vec{p}_i)\leftrightarrow (-E_f,-\vec{p}_f)$,
is {\em not} a symmetry of a nonrelativistic treatment.
\end{itemize}

   For the purpose of calculating the differential cross section, 
we make use of the Coulomb gauge, $\epsilon_0=0$, 
$\vec{q}\cdot\vec{\epsilon}=0$, $\epsilon_0'=0$, 
and $\vec{q}\,'\cdot\vec{\epsilon}\,'=0$,
and evaluate the $S$-matrix element in the laboratory frame, where
$\vec{p}_i=0$,
\begin{displaymath}
S_{fi}=-i (2\pi)^4 \delta(E_f+\omega'-E_i-\omega)
\delta^3(\vec{p}_f+\vec{q}\,'-\vec{p}_i-\vec{q})
\frac{1}{\sqrt{4\omega\omega'}(2\pi)^6}
\frac{e^2\vec{\epsilon}\cdot\vec{\epsilon}\,'^\ast}{M}.
\end{displaymath}
   Using standard techniques,\footnote{It is advantageous to discuss
these steps using box normalization instead of $\delta$-function
normalization. See Chap.\ 7.11 of Ref.\ \cite{Sakurai_85}.}
the differential cross section reads
\begin{displaymath}
d\sigma = \delta(E_f+\omega'-E_i-\omega)
\delta^3(\vec{p}_f+\vec{q}\,'-\vec{p}_i-\vec{q})
\frac{1}{\omega\omega'}
\left|\frac{e^2\vec{\epsilon}\cdot
\vec{\epsilon}\,'^\ast}{4\pi M}\right|^2 d^3q' d^3p_f.
\end{displaymath}
   After integration with respect to the momentum of the particle, we make
use of $d^3q'=\omega'^2d\omega' d\Omega$ and obtain
\begin{equation}
\label{dsdo}
\frac{d\sigma}{d\Omega}=
\left\{1-4\frac{\omega}{M}\sin^2\left(\frac{\Theta}{2}\right)
+{\cal O}\left[\left(\frac{\omega}{M}\right)^2\right]\right\}
\left|\frac{e^2\vec{\epsilon}\cdot
\vec{\epsilon}\,'^\ast}{4\pi M}\right|^2.
\end{equation}
   Averaging and summing over initial and final photon polarizations,
respectively, is easily performed by treating 
$\{\hat{q}=\hat{e}_z,\vec{\epsilon}(1)=\hat{e}_x,
\vec{\epsilon}(2)=\hat{e}_y\}$ as well as 
$\{\hat{q}',\vec{\epsilon}\,'(1),\vec{\epsilon}\,'(2)\}$
as orthonormal bases,
\begin{equation}
\label{sumpol}
\sum_{\lambda'=1}^2\left(\frac{1}{2}
\sum_{\lambda=1}^2|\vec{\epsilon}(\lambda)\cdot
\vec{\epsilon}\,'^\ast(\lambda')|^2\right)
=\frac{1}{2}[1+\cos^2(\Theta)].
\end{equation}
   Let us consider the so-called Thomson limit, {\em i.e.}, $\omega\to 0$,
for which Eq.\ (\ref{dsdo}) in combination with Eq.\ (\ref{sumpol}) reduces 
to 
\begin{displaymath}
\left.\frac{d\sigma}{d\Omega}\right|_{\omega=0}
=\frac{\alpha^2}{M^2}\frac{1+\cos^2(\Theta)}{2},
\quad \alpha=\frac{e^2}{4\pi}
\approx \frac{1}{137}.
\end{displaymath}
   The total cross section, obtained by integrating over the entire
solid angle, reproduces the classical Thomson scattering cross section 
denoted by $\sigma_T$,
\begin{equation}
\label{sigmat}
\sigma_T=\frac{8\pi}{3}\frac{\alpha^2}{M^2}.
\end{equation}
   Numerical values of the Thomson cross section for the electron, charged
pion, and the proton are shown in Table \ref{tcs}.

\subsection{Nonrelativistic Compton scattering off a composite system}
   Next we discuss Compton scattering off a composite system within 
the framework of nonrelativistic quantum mechanics.
   For the sake of simplicity, we consider a system of two particles
interacting via a central potential $V(r)$,
\begin{equation}
\label{tpp}
H_0=\frac{{\vec{p}_1}\,^2}{2m_1}+\frac{\vec{p}_2\,^2}{2m_2}
+V(|\vec{r}_1-\vec{r}_2|)
=\frac{\vec{P}\,^2}{2M}+\frac{\vec{p}\,^2}{2\mu}+V(r), 
\end{equation}
   where we introduced 
\begin{eqnarray*}
&&M=m_1+m_2,\quad
\vec{R}=\frac{m_1\vec{r}_1+m_2\vec{r}_2}{M},\quad
\vec{P}=\vec{p}_1+\vec{p}_2,
\\
&&\mu=\frac{m_1m_2}{M},\quad
\vec{r}=\vec{r}_1-\vec{r}_2,\quad 
\vec{p}=\frac{m_2\vec{p}_1-m_1\vec{p}_2}{M}.
\end{eqnarray*}
   As in the single-particle case, the electromagnetic interaction
is introduced via minimal coupling, $i\partial/\partial t\to
i\partial/\partial t-q_1\phi_1-q_2\phi_2$, 
$\vec{p}_i\to\vec{p}_i-q_i \vec{A}_i$,
resulting in the interaction Hamiltonians
\begin{eqnarray*}
H_1(t)&=&\sum_{i=1}^2\left[-\frac{q_i}{2m_i}(\vec{p}_i\cdot\vec{A_i}
+\vec{A_i}\cdot\vec{p}_i)+q_i\phi_i\right],\\
H_2(t)&=&\sum_{i=1}^2\frac{q_i^2}{2m_i}\vec{A_i}^2,
\end{eqnarray*}
where $(\phi_i,\vec{A_i})=(\phi(\vec{r}_i,t),\vec{A}(\vec{r}_i,t))$.
   In order to keep the expressions as simple as possible, we will make
some simplifying assumptions and quote the general result at the end. 
   First of all, we do not consider the spin of the constituents,
{\em i.e.}, we omit an interaction term 
\begin{displaymath}
-\sum_i\vec{\mu}_i\cdot\vec{B}_i,\quad
\vec{B}_i=\vec{\nabla}_i\times \vec{A}_i,
\end{displaymath}
   where $\vec{\mu}_i$ is an intrinsic magnetic dipole moment of the
$i$th constituent.

    Secondly, we take equal masses for the constituents, 
$m_1=m_2=m=\frac{1}{2}M$ 
and assume that one has charge $q_1=e>0$ and the second one
 is neutral, $q_2=0$.
    Finally, as a result of the gauge-invariance property
we perform the calculation within the Coulomb gauge, $\phi_i=0$.\footnote{
In actual calculations, it is highly recommended not to specify the
gauge and use gauge invariance as a check of the final result.}
   With these preliminaries, the Hamiltonian reads
\begin{displaymath}
H=H_0-\frac{e}{M}(\vec{p}_1\cdot\vec{A}_1+\vec{A}_1\cdot\vec{p}_1)
+\frac{e^2}{M}\vec{A_1}^2.
\end{displaymath}

   The $S$-matrix element is obtained in complete analogy to the previous
section within the framework of ordinary time-dependent perturbation 
theory:
\begin{equation}
\label{scp}
S_{fi}=S_{fi}^{cont}+S_{fi}^{dc}+S_{fi}^{cc},
\end{equation}
   where the seagull contribution results from the sum of the individual
contact terms and the direct-channel and crossed-channel contributions 
are more complicated as in the single-particle case, since they now also
involve excitations of the composite object.

   Using $\vec{r}_1=\vec{R}+\frac{1}{2}\vec{r}$, one obtains for 
the contact contribution 
\begin{equation}
\label{tficoncoms}
t_{fi}^{cont}=-\vec{\epsilon}\cdot\vec{\epsilon}\,'^* \frac{2e^2}{M}
\int d^3 r |\phi_0(\vec{r})|^2 
\exp\left[i(\vec{q}-\vec{q}\,')\cdot\frac{\vec{r}}{2}\right].
\end{equation}
   Since $q_2=0$, the integral is just the charge form factor 
$F[(\vec{q}-\vec{q}\,')^2]$ of the ground state,
$$ F(\vec{q}\,^2)=1-\frac{1}{6}r_E^2\vec{q}\,^2+\cdots.
$$
   We note that for a composite object, in general, the contact interactions 
of the constituents do not yet generate the complete Thomson limit.
   However, it is possible to make a unitary transformation such 
that the total Thomson amplitude is generated by a contact term
making the composite object look very similar to the point object
\cite{Jennings_87}.

   The second contribution is evaluated by inserting a complete set of states,
\begin{eqnarray}
\label{sdccc}
S_{fi}^{dc+cc}
&=&-2\pi i\delta(E_f+\omega'-E_i-\omega)\int d^3 P\sum_n\nonumber
\\
&&\times\left(
\frac{<\!\!\vec{p}_f,0|H^{em}_1|\vec{P},n\!\!><\!\!\vec{P},n| 
H^{abs}_1|\vec{p}_i,0\!\!>}{E_f+\omega'-
E_n(\vec{P})}+ 
\frac{<\!\!\vec{p}_f,0|H^{abs}_1|\vec{P},n\!\!><\!\!\vec{P},n|
H^{em}_1|\vec{p}_i,0\!\!>}{E_f-\omega-E_n(\vec{P})}\right),\nonumber\\
\end{eqnarray}
   where, in the framework of Eq.\ (\ref{tpp}), the energy of an excited 
state with intrinsic energy $\omega_n$ 
moving with c.m.\ momentum $\vec{P}$ is given by
\begin{displaymath}
E_n(\vec{P})=\frac{\vec{P}\,^2}{2M}+\omega_n.
\end{displaymath}
   In Coulomb gauge, the corresponding Hamiltonians for absorption
and emission of photons, respectively, read
\begin{displaymath}
H_1^{abs}=-\frac{2e}{M}N(\omega) \hat{\vec{p}}_1\cdot\vec{\epsilon}
\exp(i\vec{q}\cdot\vec{r}_1),\quad
H_1^{em}=-\frac{2e}{M}N(\omega') \hat{\vec{p}}_1\cdot\vec{\epsilon}\,'^\ast
\exp(-i\vec{q}\,'\cdot\vec{r}_1).
\end{displaymath}
   As in the point-object case, $S_{fi}$ is symmetric
under photon crossing   
$(\omega,\vec{q})\leftrightarrow (-\omega',-\vec{q}\,')$ 
and $\vec{\epsilon}\leftrightarrow\vec{\epsilon}\,'^\ast$.
      
   The low-energy expansion of Eq.\ (\ref{sdccc}) is obtained by expanding 
the vector potentials and the denominators in $\omega$ and $\omega'$.
   The explicit calculation is beyond the scope of the present treatment
and we will only quote the general result at the end \cite{Petrunkin_64,%
Ericson_73,Friar_75}.
   However, we find it instructive to consider the limit $\omega\to 0$:
\begin{eqnarray*}
\left.t_{fi}^{dc+cc}\right|_{\omega=0}&=&
\frac{4e^2}{M^2}\sum_n\frac{1}{\Delta\omega_n}
\left( <\!0|\vec{p}\cdot\vec{\epsilon}\,'^\ast
|n\!><\!n|\vec{p}\cdot\vec{\epsilon}|0\!>
+<\!0|\vec{p}\cdot\vec{\epsilon}
|n\!><\!n|\vec{p}\cdot\vec{\epsilon}\,'^\ast|0\!>\right),
\end{eqnarray*}  where $\Delta\omega_n=\omega_n-\omega_0$. 
   The matrix elements involve internal degrees of freedom, only.
   Making use of $\vec{p}=i\mu[H_0,\vec{r}\,]$ and applying $H_0$ 
appropriately to the right or left, the expression simplifies to 
\begin{eqnarray}
\label{tdccccpt}
\left.t_{fi}^{dc+cc}\right|_{\omega=0}&=&
-i\frac{4e^2\mu}{M^2}\sum_n \left(
<\!0|\vec{r}\cdot\vec{\epsilon}\,'^\ast
|n\!><\!n|\vec{p}\cdot\vec{\epsilon}|0\!>
-<\!0|\vec{p}\cdot\vec{\epsilon}
|n\!><\!n|\vec{r}\cdot\vec{\epsilon}\,'^\ast|0\!>\right)\nonumber\\
&=&-i\frac{4e^2\mu}{M^2}<\!0|[\vec{r}\cdot\vec{\epsilon}\,'^\ast,
\vec{p}\cdot\vec{\epsilon}]|0\!>
= \frac{e^2}{M}\vec{\epsilon}\,'^\ast\cdot\vec{\epsilon},
\end{eqnarray}
   where, again, we used the completeness relation,  
$[\vec{a}\cdot\hat{\vec{r}},\vec{b}\cdot\hat{\vec{p}}]=i\vec{a}\cdot\vec{b}$,
and $\mu=M/4$.
   Combining this result with the contact contribution of 
Eq.\ (\ref{tficoncoms}) yields the correct Thomson limit also for
a composite system.
   Indeed, it has been shown a long time ago in the more general framework 
of quantum field theory that the scattering of photons in the limit of zero
frequency is correctly described by the classical Thomson amplitude
\cite{Thirring_50,Kroll_54,Low_54,GellMann_54}. 
   We will come back to this point in the next section. 
   
   Beyond the Thomson limit, we only quote the nonrelativistic $T$-matrix 
element for Compton scattering off a spin-zero particle of mass $M$ and 
total charge $Ze$, expanded to second order in the photon energy: 
\begin{equation}
\label{tfize}
t_{fi}=\vec{\epsilon}\,'^\ast\cdot\vec{\epsilon}
\left(-\frac{(Ze)^2}{M}+4\pi\bar{\alpha}_E\omega\omega'\right)
+4\pi\bar{\beta}_M\vec{q}\,'\times\vec{\epsilon}\,'^\ast
\cdot\vec{q}\times\vec{\epsilon},
\end{equation}
where 
\begin{eqnarray}
\label{alphab}
\bar{\alpha}_E&=&\frac{\alpha Z^2 r^2_E}{3M}+2\alpha\sum_{n\neq 0}
\frac{|\!\!<\!\!n|D_z|0\!\!>\!\!|^2}{E_n-E_0},\\
\label{betab}
\bar{\beta}_M&=&-\frac{\alpha <\!\!\vec{D}\,^2\!\!>}{2M}-\frac{\alpha}{6}
<\!\!\sum_{i=1}^N \frac{q^2_i \vec{r}_i\,^2}{m_i}\!\!>+2\alpha\sum_{n\neq 0}
\frac{|\!\!<\!\!n|M_z|0\!\!>\!\!|^2}{E_n-E_0}
\end{eqnarray}
   denote the electric ($\bar{\alpha}_E$) and magnetic ($\bar{\beta}_M$)
polarizabilities of the system.
   In these equations  
\begin{displaymath}
\vec{D}=\sum_{i=1}^N q_i(\vec{r}_i-\vec{R})
\end{displaymath}
   refers to the intrinsic electric dipole operator
and 
\begin{displaymath} 
\vec{M}=
\sum_{i=1}^N
\left[
\frac{\hat{q}_i}{2m_i}(\vec{r}_i-\vec{R})\times
(\vec{p}_i-\frac{m_i}{M}\vec{P})
+\vec{\mu}_i\right]
\end{displaymath}
   to the magnetic dipole operator, where the possibility of magnetic moments 
of the constituents has now been included.
   The electromagnetic polarizabilities describe the response of the internal
degrees of freedom of a system to a small external electromagnetic 
perturbation.
   For example, in atomic physics the second term of Eq.\ (\ref{alphab})
is related to the quadratic Stark effect describing the energy shift
of an atom placed in an external electric field. 
   We will come back to an interpretation of the electric polarizability
in terms of a classical analogue in the next subsection.
   
   Finally, let us discuss the influence of the electromagnetic 
polarizabilities on the differential Compton-scattering cross section.
   We restrict ourselves to the leading term due to the interference of 
the Thomson amplitude with the polarizability contribution.
   The evaluation of that term requires, in addition to Eq.\ (\ref{sumpol}),
the sum
\begin{displaymath}
\sum_{\lambda,\lambda'}Re\{\vec{\epsilon}\,'^\ast\cdot\vec{\epsilon}
\hat{q}\,'\times\vec{\epsilon}\,'\cdot\hat{q}\times\vec{\epsilon}\,^\ast\}
=2\cos(\Theta),
\end{displaymath}
and one obtains
\begin{eqnarray}
\label{dsdopol}
\frac{d\sigma}{d\Omega}&=&
\left[1-4\frac{\omega}{M}\sin^2\left(\frac{\Theta}{2}\right)
+{\cal O}\left(\frac{\omega^2}{M^2}\right)\right]
\left\{\frac{1}{2}[1+\cos^2(\Theta)]
\frac{\alpha^2 Z^4}{M^2}\right.\nonumber\\
&&\left. -[1+\cos^2(\Theta)]\frac{\alpha Z^2}{M}\bar{\alpha}_E \omega\omega'
-2\cos(\Theta)\frac{\alpha Z^2}{M}\bar{\beta}_M\omega\omega'
+{\cal O}(\omega^2\omega'^2)\right\}.
\end{eqnarray}
   The differential cross sections at $\Theta=0^\circ,90^\circ$, and
$180^\circ$ are sensitive to $\bar{\alpha}_E+\bar{\beta}_M$,
$\bar{\alpha}_E$, and $\bar{\alpha}_E-\bar{\beta}_M$, respectively.

\subsection{Classical interpretation}
   The prototype example for illustrating the concept of an electric 
polarizability is a system of two harmonically bound point particles of 
mass $m$ with opposite charges $+q$ and $-q$ 
\cite{Friar_89,Holstein_90}:
\begin{displaymath}
H_0=\frac{\vec{p}_1\,^2}{2m}+\frac{\vec{p}_2\,^2}{2m}+\frac{\mu\omega^2_0}{2}
\vec{r}\,^2,\quad\mu=\frac{m}{2},\quad \vec{r}=\vec{r}_1-\vec{r}_2,
\end{displaymath}
   where we neglect the Coulomb interaction between the charges.
   If a static, uniform, external electric field 
$$\vec{E}=E_0\hat{e}_z$$
   is applied to this system, the equilibrium position is determined by
\begin{displaymath}
\mu\ddot{z}=-\mu\omega_0^2z+qE_0 \stackrel{!}{=}0, 
\end{displaymath}
leading to 
   $$z_0=\frac{qE_0}{\mu\omega_0^2}.$$
   The electric polarizability $\alpha_E$ is defined via the relation between
the induced electric dipole moment and the external field\footnote{The 
factor $4\pi$ results from our use of the
Heaviside-Lorentz units instead of the Gaussian system.} 
\begin{displaymath}
\vec{p}=qz_0\hat{e}_z\equiv 4\pi\alpha_E \vec{E}.
\end{displaymath}
   For the harmonically bound system, $\alpha_E$ is
proportional to the inverse of the spring constant,   
\begin{displaymath}
\alpha_E=\frac{q^2}{4\pi}\frac{1}{\mu\omega^2_0},
\end{displaymath}
{\em i.e.}, it is a measure of the stiffness or
rigidity of the system \cite{Holstein_90}.
   The potential energy associated with the induced electric dipole moment
reads 
\begin{equation}
\label{valpha}
V=-2\pi\alpha_E\vec{E}^2=-\frac{1}{2}\vec{p}\cdot\vec{E},
\end{equation}
   where the factor $\frac{1}{2}$ results from the interaction
of an induced rather than a permanent electric dipole moment with 
the external field.
   Similarly, the potential of an induced magnetic dipole,
$\vec{m}=4\pi\beta_M \vec{H}$, is given by 
\begin{equation}
\label{vbeta}
V=-2\pi\beta_M \vec{H}^2.
\end{equation}

\section{Compton scattering in quantum field theory}
   Now that we have discussed Compton scattering in the framework of
{\em nonrelativistic} quantum mechanics, we will turn to a description 
in the context of quantum field theory. 
   Generally, we will consider the case of the nucleon but will restrict
ourselves to the pion whenever this allows for a (substantial) simplification
without loss of generality.
   We will direct our attention to the influence of hadron structure on the 
description of electromagnetic processes.
   In particular, we will emphasize the power of Ward-Takahashi identities
\cite{Ward_50,Takahashi_57}.
   First of all, we will describe the simplest electromagnetic vertex, 
namely, the interaction of a single photon with a charged pion.
   Using the method of Gell-Mann and Goldberger \cite{GellMann_54},
we will derive the low-energy and low-momentum behavior of the (virtual)
Compton-scattering tensor based upon Lorentz invariance, gauge invariance,
crossing symmetry, and the discrete symmetries. 
   Finally, we will consider Compton scattering off a pion to illustrate
why off-shell effects are not directly observable.

\subsection{Electromagnetic vertex of a charged pion}
   For the purpose of illustrating the power of symmetry considerations,
we explicitly discuss the most general electromagnetic vertex of an
off-shell pion.
   We will formally introduce the concept of {\em form 
functions} by parameterizing the electromagnetic three-point Green's function
of a pion.
   In this context, we distinguish between {\em form factors} and 
{\em form functions}, the former representing observables, which is, 
in general, not true for form functions.

   Let us define the three-point Green's function of two unrenormalized 
pion field operators $\pi^+(x)$ and $\pi^-(y)$ and the electromagnetic 
current operator $J^\mu(z)$ as\footnote{$\pi^{+/-}(x)$ 
destroys a $\pi^{+/-}$ or creates a $\pi^{-/+}$.}
\begin{equation}
\label{tpgf}
G^\mu(x,y,z)=<\!\! 0|T\left[\pi^+(x) \pi^-(y)J^{\mu}(z)\right]|0\!\!>,
\end{equation}
and consider the corresponding momentum-space Green's function 
\begin{equation}
\label{gmu}
(2\pi)^4 \delta^4(p_f-p_i-q) G^{\mu}(p_f,p_i)=
\int d^4x\, d^4y\, d^4z\, e^{i(p_f \cdot x - p_i \cdot y-q\cdot z )}
G^\mu(x,y,z),
\end{equation}
where $p_i$ and $p_f$ are the four-momenta corresponding to the pion lines
entering and leaving the vertex, respectively, and $q=p_f-p_i$ is the 
momentum transfer at the vertex.
   Defining the renormalized three-point Green's function $G^\mu_R$ as
\begin{equation}
\label{gmur}
G^{\mu}_R(p_f,p_i) = Z^{-1}_{\phi} Z^{-1}_J G^{\mu}(p_f,p_i),
\end{equation}
where $Z_{\phi}$ and $Z_J$ are renormalization 
constants,\footnote{In fact, $Z_J=1$ due to gauge invariance.}
   we obtain the one-particle irreducible, renormalized three-point Green's 
function by removing the propagators at the external lines,
\begin{equation}
\label{gammamuirr}
\Gamma^{\mu,irr}_R(p_f,p_i) =
[i \Delta_R(p_f)]^{-1} G^{\mu}_R(p_f,p_i)[i\Delta_R(p_i)]^{-1},
\end{equation}
where $\Delta_R(p)$ is the full, renormalized propagator.
   From a perturbative point of view, $\Gamma^{\mu,irr}_R$ is made up of 
those Feynman diagrams which cannot be disconnected by cutting any one single 
internal line.

   In the following we will discuss a few model-independent properties of 
$\Gamma^{\mu,irr}_R(p_f,p_i)$.
\begin{enumerate}
\item Imposing Lorentz covariance, the most general parameterization of 
$\Gamma^{\mu,irr}_R$ can be written in terms of two independent four-momenta,
$P^\mu=p_f^\mu+p_i^\mu$ and $q^\mu=p_f^\mu-p_i^\mu$, respectively, 
multiplied by Lorentz-scalar form functions $F$ and $G$ depending on 
three scalars, {\em e.g.}, $q^2$, $p_i^2$, and $p_f^2$,
\begin{equation}
\label{par}
\Gamma^{\mu,irr}_R(p_f,p_i) =
(p_f+p_i)^{\mu} F(q^2,p_f^2,p_i^2) +
(p_f-p_i)^{\mu} G(q^2,p_f^2,p_i^2).
\end{equation}
\item Time-reversal symmetry results in 
\begin{equation}
\label{trs}
F(q^2,p_f^2,p_i^2)=F(q^2,p_i^2,p_f^2), \quad 
G(q^2,p_f^2,p_i^2)=-G(q^2,p_i^2,p_f^2).
\end{equation}
   In particular, from Eq.\ (\ref{trs}) we conclude that 
$G(q^2,M^2_\pi,M^2_\pi)=0$.
   This, of course, corresponds to the well-known fact that a spin-0 particle 
has only one electromagnetic form factor, $F(q^2)$. 
\item Using the charge-conjugation properties $J^\mu\mapsto-J^\mu$ and
$\pi^+\leftrightarrow\pi^-$, it is straightforward to see that form 
functions of particles are just the negative of form functions of 
antiparticles. 
   In particular, the $\pi^0$ does not have any electromagnetic form 
functions even off shell, since it is its own antiparticle.
\item Due to the hermiticity of the electromagnetic current operator,
$F(q^2)$ is real in the spacelike region $q^2\leq 0$:
\begin{eqnarray*}
(p_f+p_i)^\mu F^\ast(q^2)&=&<p_f|J^\mu(0)|p_i>^\ast
=<p_i|{J^\mu}^\dagger(0)|p_f>
=<p_i|J^\mu(0)|p_f>\\
&=&(p_i+p_f)^\mu F(q^2)\,\,\mbox{for}\,\, q^2\leq 0.
\end{eqnarray*}
   \item After writing out the various time orderings in Eq.\ (\ref{tpgf}),
let us consider the divergence 
\begin{eqnarray}
\label{wtstart}
\partial_\mu^z G^\mu(x,y,z)&=&
<\!\!0|T[\pi^+(x)\pi^-(y)\partial_\mu J^\mu(z)]|0\!\!>\nonumber\\
&&+\delta(z^0-x^0)<\!\!0|T\{[J^0(z),\pi^+(x)]\pi^-(y)\}|0\!\!>\nonumber\\
&&+\delta(z^0-y^0)<\!\!0|T\{\pi^+(x)[J^0(z),\pi^-(y)]\}|0\!\!>.
\end{eqnarray}
   Current conservation at the operator level, 
$\partial_\mu J^\mu(z)=0$, together with the equal-time 
commutation relations of 
the electromagnetic charge-density operator with the pion field 
operators,\footnote{
Note that both equations are related by taking the adjoint.}
\begin{eqnarray}
\label{comrel}
[J^0(x),\pi^-(y)] \delta (x^0-y^0) & = & \delta^4(x-y) \pi^-(y), \nonumber \\
{[}J^0(x),\pi^+(y)]  \delta (x^0-y^0) & = & -\delta^4(x-y) \pi^+(y),
\end{eqnarray}
are the basic ingredients for obtaining Ward-Takahashi identities 
\cite{Ward_50,Takahashi_57} for electromagnetic processes.
   For example, we obtain from Eq.\ (\ref{wtstart}) 
\begin{equation}
\label{partialg}
\partial_\mu^z G^\mu(x,y,z)=\left[\delta^4(z-y)-\delta^4(z-x)\right]
<\!\!0|T[\pi^+(x)\pi^-(y)]|0\!\!>.
\end{equation}
   Taking the Fourier transformation of Eq.\ (\ref{partialg}), peforming
a partial integration, and repeating the same steps which lead from
Eq.\ (\ref{gmu}) to (\ref{gammamuirr}), one obtains the celebrated 
Ward-Takahashi identity for the electromagnetic vertex 
\begin{equation}
\label{wti}
q_{\mu} \Gamma^{\mu,irr}_R(p_f,p_i) =
\Delta_R^{-1}(p_f)-\Delta_R^{-1}(p_i).
\end{equation}
   In general, this technique can be applied to obtain Ward-Takahashi 
identities relating Green's functions which differ by insertions of the 
electromagnetic current operator.

   Inserting the parameterization of the irreducible vertex, Eq.\ (\ref{par}),
into the Ward-Takahashi identity, Eq.\ (\ref{wti}), the form functions
$F$ and $G$ are constrained to satisfy
\begin{equation}
\label{constraint}
(p_f^2-p_i^2) F(q^2,p_f^2,p_i^2)+q^2 G(q^2,p_f^2,p_i^2)
= \Delta^{-1}_R(p_f)-\Delta^{-1}_R(p_i).
\end{equation}
   From Eq.\ (\ref{constraint}) it can be shown that, given a consistent 
calculation of $F$, the propagator of the particle, $\Delta_R$, as well as 
the form function $G$ are completely determined (see Appendix A of 
Ref.\ \cite{Rudy_94} for details).
   The Ward-Takahashi identity thus provides an important consistency check 
for microscopic calculations.
\item As the simplest example, one may consider a structureless
``point pion'':
\begin{displaymath}
\Gamma^\mu(p_f,p_i)=(p_f+p_i)^\mu,\quad 
q_\mu \Gamma^\mu=p_f^2-p_i^2=(p_f^2-m_\pi^2)-(p^2_i-m_\pi^2),
\end{displaymath}
{\em i.e.}, $F(q^2,p_f^2,p_i^2)=1$, $G(q^2,p_f^2,p_i^2)=0$.
\item As was already pointed out in Ref.\ \cite{Barton_65}, use of 
\begin{displaymath}
\Gamma^\mu(p_f,p_i)=(p_f+p_i)^\mu F(q^2)
\end{displaymath}
leads to an inconsistency, since the left-hand side of the 
corresponding Ward-Takahashi identity depends on $q^2$,
 whereas the right-hand side only depends on $p_f^2$ and $p^2_i$.
\end{enumerate}
   The nucleon case is more complicated due to spin and the most
general form of the irreducible, electromagnetic vertex can be
expressed in terms of 12 operators and associated form functions.
   The interested reader is referred to Refs.\ \cite{Bincer_60,Naus_87}.

   Finally, it is important to emphasize that the off-shell behavior of form 
functions is representation dependent, {\em i.e.}, form functions are, in 
general, not observable.
   In the context of a Lagrangian formulation, this can be understood
as a result of field transformations \cite{Chisholm_61,%
Kamefuchi_61,Coleman_69,Scherer_95a}.
   This does not render the previous discussion useless, rather the 
Ward-Takahashi identities provide important consistency relations between the 
building blocks of a quantum-field-theoretical description.

\subsection{Low-energy theorem for the Compton-scattering tensor}
   The Compton-scattering tensor $V^{\mu\nu}_{s_is_f}$ is defined 
through a Fourier transformation of the time-ordered product of two
electromagnetic current operators evaluated between on-shell
nucleon states:\footnote{In the following, we will consider the proton
case.}
\begin{eqnarray}
\label{cst}
\lefteqn{(2\pi)^4\delta^4(p_f+q'-p_i-q) V^{\mu\nu}_{s_i s_f}(p_f,q';p_i,q)=}
\nonumber\\
&&\int d^4x d^4y e^{-iq\cdot x}e^{iq'\cdot y}
<\!N(p_f,s_f)|T[J^\mu(x)J^\nu(y)]|N(p_i,s_i)\!>.
\end{eqnarray}
   The relation to the invariant amplitude of real Compton scattering 
(RCS),\footnote{We use the convention of Bjorken and Drell 
\cite{Bjorken_64}.} is given by 
\begin{equation}
\label{mrcs}
{\cal M}=-e^2\epsilon_\mu\epsilon'^\ast_\nu
\left. V^{\mu\nu}_{s_i s_f}(p_f,q';p_i,q)\right|_{q^2=q'^2=0}.
\end{equation}
   In RCS, $V^{\mu\nu}_{s_is_f}$ can only be tested for a rather 
restricted range of variables $q^\mu$ and $q'^\mu$ and, furthermore, only 
the transverse components of  $V^{\mu\nu}_{s_is_f}$ are accessible.
   The expression ``virtual Compton scattering'' (VCS) refers to the 
situation, where one or both photons are virtual.
   We will primarily be concerned with the case $q^2<0$ and $q'^2=0$ which,
{\em e.g.}, for the proton can be tested in the reaction $e^-p\to e^-p\gamma$.

   In the following, we will discuss the low-energy and small-momentum 
behavior of the Compton-scattering tensor. 
   Our discussion will closely follow the method of Gell-Mann and
Goldberger \cite{GellMann_54,Kazes_59}.
     Let us first recall the distinction between 
$V^{\mu\nu}_{s_is_f}$ and $\Gamma^{\mu\nu}$,\footnote{We omit the superscript
{\em irr} and the subscript $R$, respectively.}
\begin{displaymath}
V^{\mu\nu}_{s_is_f}(p_f,q';p_i,q)=
\bar{u}(p_f,s_f)\Gamma^{\mu\nu}(P,q',q)u(p_i,s_i),
\end{displaymath}
where $P=p_f+p_i$. 
   In $V^{\mu\nu}_{s_is_f}$ the nucleon is assumed to be on shell,  
$p_i^2=p_f^2=M^2$, whereas $\Gamma^{\mu\nu}$ is defined for arbitrary 
$p_i^2$ and $p_f^2$, {\em i.e.}, $\Gamma^{\mu\nu}$ is the analogue of
$\Gamma^{\mu,irr}_R$ of Eq.\ (\ref{gammamuirr}).
   We divide the contributions to $\Gamma^{\mu\nu}$ into two classes, 
$A$ and $B$,
\begin{displaymath}
\Gamma^{\mu\nu}=\Gamma^{\mu\nu}_A+\Gamma^{\mu\nu}_B,
\end{displaymath}
where class $A$ consists of the s- and u-channel pole terms and class $B$ 
contains all the other contributions.
   With this separation all terms which are irregular for 
$q^\mu\rightarrow 0$ (or $q'^\mu\rightarrow 0$) are contained in
class $A$, whereas class $B$ is regular in this limit.
   Strictly speaking, one also assumes that there are no massless particles in
the theory which could make a low-energy expansion in terms of kinematical
variables impossible \cite{Low_54}.
   The contribution due to t-channel exchanges, such as a
$\pi^0$, is taken to be part of class $B$.
 
   We express class $A$ in terms of the full renormalized propagator and 
the irreducible electromagnetic vertices,
\begin{equation}
\label{gammaa}
\Gamma^{\mu\nu}_A=\Gamma^\nu(p_f,p_f+q')iS(p_i+q)
\Gamma^\mu(p_i+q,p_i)+\Gamma^\mu(p_f,p_f-q)iS(p_i-q') \Gamma^\nu(p_i-q',p_i).
\end{equation}
   Note that $\Gamma^{\mu\nu}_A$ is symmetric under photon crossing,
$q\leftrightarrow -q'$ and $\mu\leftrightarrow\nu$, {\em i.e.},
$\Gamma_A^{\mu\nu}(P,q,q')=\Gamma_A^{\nu\mu}(P,-q',-q)$.
   Since this is also the case for the total $\Gamma^{\mu\nu}$,
class $B$ must be separately crossing symmetric \cite{GellMann_54}.
   In analogy to the previous section, Ward-Takahashi identities
can be obtained for $\Gamma^\mu$ and $\Gamma^{\mu\nu}$,
\begin{eqnarray}
\label{wtn1}
q_\mu \Gamma^\mu(p_f,p_i)&=&S^{-1}(p_f)-S^{-1}(p_i),\\
\label{wtn2}
q_\mu \Gamma^{\mu\nu}(P,q',q)&=&
i\left[S^{-1}(p_f)S(p_f-q)\Gamma^\nu(p_f-q,p_i)
-\Gamma^\nu(p_f,p_i+q)S(p_i+q)S^{-1}(p_i)\right].
\end{eqnarray}
   Using Eq.\ (\ref{wtn1}), one obtains the following constraint 
for class $A$ as imposed by gauge invariance:
\begin{eqnarray}
\label{cgigammaa}
q_\mu \Gamma^{\mu\nu}_A(P,q,q')&=&i\left[\Gamma^\nu (p_f,p_f+q')
-\Gamma^\nu(p_i-q',p_i)
+S^{-1}(p_f)S(p_i-q')\Gamma^\nu(p_i-q',p_i)\right.\nonumber\\
&&\left.
-\Gamma^\nu(p_f,p_f+q')S(p_i+q)S^{-1}(p_i)\right].
\end{eqnarray}
   Eqs.\ (\ref{wtn2}) and (\ref{cgigammaa}) can now be combined
to obtain a constraint for class $B$
\begin{equation}
\label{constraintb}
q_\mu \Gamma^{\mu\nu}_B=q_\mu(\Gamma^{\mu\nu}-\Gamma^{\mu\nu}_A)
=i[\Gamma^\nu(p_i-q',p_i)-\Gamma^\nu(p_f,p_f+q')].
\end{equation}
   At this point, we make use of the freedom of choosing a convenient 
representation for $\Gamma^\mu$ below the pion production threshold,
\begin{equation}
\label{gammaeff}
\Gamma^\mu_{\mbox{\footnotesize eff}}(p_f,p_i)
=\gamma^\mu F_1(q^2)+i\frac{\sigma^{\mu\nu}q_\nu}{2M}
F_2(q^2)
+q^\mu q\hspace{-.5em}/\hspace{.5em} \frac{1-F_1(q^2)}{q^2},
\end{equation}
where $F_1$ and $F_2$ are the Dirac and Pauli form factors of the
proton, respectively.
   The fundamental reason for this assumption is the fact that one
can perform transformations of the fields in an effective Lagrangian
which do not change the physical observables but which allow to 
a certain extent to transform away the off-shell dependence at the
vertices.
   We will come back to this point in the next section.

   Eq.\ (\ref{gammaeff}) contains on-shell quantities only, and 
satisfies the Ward-Takahashi identity in combination 
with the free Feynman propagator, 
$$
q_\mu\Gamma^\mu_{\mbox{\footnotesize eff}}=q\hspace{-.5em}/\hspace{.5em}
=S^{-1}_F(p_f)-S^{-1}_F(p_i).
$$
   In this representation the constraint for class $B$ is particularly
simple:
\begin{equation}
\label{constraintbs}
q_\mu \Gamma^{\mu\nu}_B=0.
\end{equation}
   In order to solve this equation, one first makes an ansatz for class $B$,
\begin{equation}
\label{gammabansatz}
$$\Gamma^{\mu\nu}_B(P,q',q)=a^{\mu,\nu}(P,q')+b^{\mu\rho,\nu}(P,q')q_\rho
+\cdots
\end{equation}
which is inserted into Eq.\ (\ref{constraintbs}),
\begin{equation}
\label{consb3}
0=a^{\mu,\nu}(P,q')q_\mu  
+b^{\mu\rho,\nu}(P,q')q_\mu q_\rho+
\cdots.
\end{equation}
   The constraints due to crossing symmetry, and the discrete symmetries
are imposed and Eq.\ (\ref{consb3}) is
solved as a power series in $q$ and $q'$.
   At this point, off-shell kinematics is required in order to
be able to treat $q$, $q'$, and $P$ as completely independent. 
   For example, using off-shell kinematics 6 invariant scalars can be
formed from $q$, $q'$, and $P$, whereas for the on-shell case,
$p_i^2=p_f^2=M^2$, only four of these combinations are independent.
   A detailed description of this procedure can be found in 
Ref.\ \cite{Scherer_96} and we will only summarize the key results.
   Class $B$ contains no constant terms and no terms at 
${\cal O}(q)$ or ${\cal O}(q')$.
   Combining this observation with an evaluation of class $A$ 
generates, for RCS, the LET of Refs.\ \cite{Low_54,GellMann_54}.
   At ${\cal O}(2)$ one finds two terms which can be related to the 
electromagnetic polarizabilities $\bar{\alpha}_E$ and $\bar{\beta}_M$.
   The framework is general enough to also account for virtual photons.
   In particular, no new polarizability appears in the longitudinal part.
   In fact, this result has also been obtained in the framework of
effective Lagrangians in Ref.\ \cite{Lvov_93}.

   The result for the RCS amplitude can be summarized as
\begin{eqnarray}
\label{mresrcs}
{\cal M}&=&-ie^2\bar{u}(p_f,s_f)
\left[\epsilon'^\ast\cdot\Gamma(-q') S_F(p_i+q)\epsilon\cdot\Gamma(q)
+\epsilon\cdot\Gamma(q) S_F(p_i-q')\epsilon'^\ast\cdot\Gamma(-q')\right.
\nonumber\\
&&-\frac{4\pi}{e^2}\bar{\beta}_M
(\epsilon\cdot\epsilon'^\ast q\cdot q'
-\epsilon\cdot q'\epsilon'^\ast\cdot q)
+\frac{\pi}{e^2 M^2}(\bar{\alpha}_E+\bar{\beta}_M)
(\epsilon\cdot\epsilon'^\ast P\cdot q P\cdot q'
+\epsilon\cdot P\epsilon'^\ast\cdot Pq\cdot q'\nonumber\\
&&\left.
-\epsilon\cdot P\epsilon'^\ast\cdot q P\cdot q'
-\epsilon\cdot q'\epsilon'^\ast\cdot P P\cdot q)
+{\cal O}(3)\right]u(p_i,s_i),
\end{eqnarray}
   where we introduced the abbreviation 
\begin{equation}
\label{gmureal}
\Gamma^\mu(q)=\gamma^\mu +i\frac{\sigma^{\mu\nu}q_\nu}{2M}\kappa,
\quad \kappa=1.79.
\end{equation}
   Here, the electromagnetic polarizabilities are defined with respect to
``Born terms'' which have been calculated with the vertices of Eqs.\ 
(\ref{gammaeff}) or (\ref{gmureal}) for RCS.
   In particular, with such a choice the Born terms are separately
gauge invariant.
   As a matter of fact, this is not always the case, since, in principle, one 
could have started to calculate the Born terms with on-shell equivalent 
electromagnetic vertices containing the Sachs form factors $G_E$ and $G_M$ or 
the Barnes form factors $H_1$ and $H_2$.
   Then class $B$ would have taken a different form even though the
final result for the total amplitude, of course, has to be the same.
    For a more detailed discussion of the ambiguity of defining
``Born terms,'' see Sec.\ IV of 
Ref.\ \cite{Scherer_96} as well as Ref.\ \cite{Fearing_98}.
   
   Table \ref{rcspoln} contains a selection of results of various models for 
the electromagnetic polarizabilities which have to be compared with the 
empirical numbers of Tables \ref{rcspolpemp} and \ref{rcspolnemp}.
   Within the framework of an effective Lagrangian it was shown 
in Ref.\ \cite{Lvov_93} that, in a covariant approach, the Compton 
polarizabilities $\bar{\alpha}_E$ and $\bar{\beta}_M$ coincide with the 
parameters determining the energy shifts in Eqs.\ (\ref{valpha}) and
(\ref{vbeta}).  
   This should be compared with a nonrelativistic treatment, where, say,
in the quadratic Stark effect only the second term of Eq.\ (\ref{alphab})
appears in the energy shift.
      Whenever comparing different results, the original references should
be consulted in order to see whether the predictions have been obtained 
in a nonrelativistic or a covariant framework.

   The sum of the electric and magnetic polarizabilities is constrained by 
the Baldin sum rule \cite{Baldin_60},
\begin{equation}
\label{baldinsumrule}
(\bar{\alpha}_E+\bar{\beta}_M)_N=\frac{1}{2\pi^2}\int_{\omega_{thr}}^\infty
\frac{\sigma^{\mbox{\footnotesize tot}}_N(\omega)}{
\omega^2}d\omega,
\end{equation}
where $\sigma^{\mbox{\footnotesize tot}}_N(\omega)$ is the total 
photoabsorption cross section.
   Eq.\ (\ref{baldinsumrule}) is obtained via a once-subtracted 
dispersion relation for the spin-averaged forward Compton amplitude
using the optical theorem together with the LET.
   An evaluation of the integral requires an extrapolation of available data 
to infinity (the results are given in units of $10^{-4}$ fm$^3$),
\begin{eqnarray}
\label{baldin}
(\bar{\alpha}_E+\bar{\beta}_M)_p&=&14.2 \pm 0.3,\quad\cite{Damashek_70}
\nonumber\\
(\bar{\alpha}_E+\bar{\beta}_M)_n&=& 15.8 \pm 0.5,\quad\cite{Lvov_79}
\nonumber\\
(\bar{\alpha}_E+\bar{\beta}_M)_p&=&13.69\pm 0.14,\quad\cite{Babusci_98a}
\nonumber\\
(\bar{\alpha}_E+\bar{\beta}_M)_n&=&14.40\pm 0.66,\quad\cite{Babusci_98a}
\end{eqnarray}
   where the last two results correspond to the most recent analysis.

   Finally, we mention that four spin polarizabilities $\gamma_i$ parameterize
the amplitude at ${\cal O}(3)$ \cite{Ragusa_93}.
   These spin-dependent terms have recently received considerable attention
but a discussion of these structure constants is beyond
the scope of the present treatment and we refer the interested reader to 
Refs.\ \cite{Hemmert_98,Drechsel_98,Tonnison_98,Babusci_98b}.

\subsection{Compton scattering and off-shell effects}
   The issue of how to treat particles with ``internal'' structure 
as soon as they do not satisfy on-mass-shell kinematics has a long history.
   As an example, we have seen in Sec.\ III.A that the electromagnetic vertex 
of a pion involving off-mass-shell momenta is more complicated than for 
asymptotically free states.
   It is therefore natural to ask how such off-shell effects show up in
observables and, in particular, whether they can be extracted from
empirical information.
   Several attempts have been made to calculate off-shell effects 
within microscopic models and to estimate their importance in 
physical observables (see, {\em e.g.}, Refs.\ 
\cite{Bincer_60,Naus_87,Nyman_70,Tiemeijer_90,Kondratyuk_98}).
 
   We will argue in this section that off-shell effects are not 
only model dependent but also representation dependent and thus
not directly measurable.
   In studying off-shell effects, we find that nucleon spin is an inessential
complication.
   We use Compton scattering off a pion in the framework of chiral 
perturbation
theory (ChPT) only as a {\em vehicle} to illustrate the point we want to make.
   Our conclusions are more general, {\em i.e.}, apply to other processes
as well, and do not rely on chiral symmetry.
   We choose ChPT, since it provides a complete and consistent 
field-theoretical framework.

\subsubsection{The chiral Lagrangian and field redefinitions}
   In this section we shall give a brief introduction to those aspects of
chiral perturbation theory \cite{Weinberg_79,Gasser_84,Gasser_85} 
which are relevant for a discussion of off-shell Green's functions.
   We will introduce the concept of field transformations since it turns out 
to be important for interpreting the meaning of form functions.
  
   In the limit of massless $u$, $d$, and $s$ quarks, the $QCD$ Lagrangian
exhibits a chiral $\mbox{SU(3)}_L\times\mbox{SU(3)}_R$ symmetry which is
assumed to be spontaneously broken to a subgroup isomorphic to
$\mbox{SU(3)}_V$, giving rise to eight massless pseudoscalar Goldstone 
bosons with vanishing interactions in the limit of zero energies.
   In ChPT the chiral symmetry is mapped onto the most general effective 
Lagrangian for the interaction of these Goldstone bosons.
   The corresponding Lagrangian is organized in a momentum 
expansion \cite{Gasser_84,Gasser_85,Fearing_96},
\begin{equation}  
\label{leff}
  {\cal L}_{\mbox{\footnotesize eff}}
={\cal L}_2+{\cal L}_4+ {\cal L}_6 +\cdots, 
\end{equation}
   where the index $2n$ denotes $2n$ derivatives.
   Couplings to external fields, such as the electromagnetic
field, as well as explicit symmetry breaking due to the finite
quark masses, can be systematically incorporated into the effective
Lagrangian.
   Weinberg's power counting scheme allows for a classification
of Feynman diagrams by establishing a relation between the momentum
expansion and the loop expansion.
   Thus, a perturbative scheme is set up in terms of external momenta 
which are small compared to some scale.
   Covariant derivatives and quark-mass terms are counted as ${\cal O}(p)$ 
and ${\cal O}(p^2)$, respectively, in the power counting scheme.

   The most general chiral Lagrangian at ${\cal O}(p^2)$ is given by
\begin{equation}
\label{l2}
{\cal L}_2 = \frac{F_0^2}{4} \mbox{Tr} \left[ D_{\mu} U (D^{\mu}U)^{\dagger} 
+\chi U^{\dagger}+ U \chi^{\dagger} \right],\quad 
U(x)=\exp\left( i\frac{\phi(x)}{F_0} \right ),
\end{equation}
where 
\begin{equation}
\label{phi}
\phi(x)=\left ( 
\begin{array}{ccc}
\pi^0+\frac{1}{\sqrt{3}}\eta & \sqrt{2} \pi^+ & \sqrt{2} K^+ \\
\sqrt{2} \pi^- & -\pi^0+\frac{1}{\sqrt{3}}\eta & \sqrt{2} K^0 \\
\sqrt{2} K^- & \sqrt{2} \bar{K}^0 & -\frac{2}{\sqrt{3}}\eta
\end{array}
\right ).
\end{equation}
   The quark-mass matrix is contained in $\chi=2 B_0\, \mbox{diag}
(m_u,m_d,m_s)$.
$B_0$ is related to the quark condensate $<\!\!\bar{q}q\!\!>$, 
$F_0\approx 93$ MeV denotes the pion-decay constant in the chiral 
limit.
   The covariant derivative $D_\mu U = \partial_\mu U +ie A_\mu [Q,U]$,
where $Q=\mbox{diag}(2/3,-1/3,-1/3)$ is the quark-charge matrix, $e>0$, 
generates a coupling to the electromagnetic field $A_\mu$.
   Finally, the equation of motion (EOM) obtained from ${\cal L}_2$ reads 
\begin{equation}
\label{eom2}
{\cal O}^{(2)}_{EOM}(U)=D^2 U U^\dagger - U (D^2 U)^\dagger 
-\chi U^\dagger +U \chi^\dagger
+\frac{1}{3}\mbox{Tr}\left(\chi U^\dagger-U\chi^\dagger\right)=0.
\end{equation}
   The most general structure of ${\cal L}_4$ was first written down by 
Gasser and Leutwyler (see Eq.\ (6.16) of Ref.\ \cite{Gasser_85}), 
\begin{equation}
\label{l4}
{\cal L}_4=L_1\left\{\mbox{Tr}[D_\mu U (D^\mu U)^\dagger]\right\}^2 + \cdots,
\end{equation}
and introduces 10 physically relevant low-energy coupling constants $L_i$.
   
   We now discuss the concept of field transformations 
\cite{Chisholm_61,Kamefuchi_61,Coleman_69,Scherer_95a}
by introducing a field redefinition,
\begin{equation}
\label{up}
U'=\mbox{exp}(iS)U=U+iSU+\cdots,
\end{equation}
where $S=S^\dagger$ and $\mbox{Tr}(S)=0$, and then look for generators $S$
which i) are of ${\cal O}(p^2)$, ii) guarantee that $U'$ has the
correct $\mbox{SU(3)}_L\times\mbox{SU(3)}_R$ transformation properties,
iii) produce the correct parity and charge-conjugation behavior,
$P:U'(\vec{x},t)\mapsto U'^\dagger(-\vec{x},t)$,
$C:U'\mapsto U'^T$.
   After some algebra (see Ref.\ \cite{Scherer_95a} for details) one
finds two such generators at ${\cal O}(p^2)$,
\begin{equation}
\label{sgen}
S=i\alpha_1[D^2 U U^\dagger - U (D^2U)^\dagger]
+i\alpha_2[\chi U^\dagger- U\chi^\dagger
-\frac{1}{3}\mbox{Tr}(\chi U^\dagger - U\chi^\dagger)],
\end{equation}
   where $\alpha_1$ and $\alpha_2$ are arbitrary real parameters with 
dimension $energy^{-2}$.
    If we insert $U'$ into ${\cal L}_{\mbox{\footnotesize eff}}$ 
of Eq.\ (\ref{leff}), we obtain 
\begin{equation}
\label{lagup}
{\cal L}_{\mbox{\footnotesize eff}}
(U)\mapsto{\cal L}_{\mbox{\footnotesize eff}}(U')
={\cal L}_2(U)+\Delta {\cal L}_2(U) 
+{\cal L}_4(U) + {\cal O}(p^6),
\end{equation}
where, to leading order in $S$, $\Delta{\cal L}_2(U)$ is given by
\begin{equation}
\label{dlag2}
\Delta {\cal L}_2(U)= \mbox{total divergence} +\frac{F^2_0}{4}
\mbox{Tr}(iS {\cal O}_{EOM}^{(2)}) + {\cal O}(p^6).
\end{equation}
   As usual, the total divergence is irrelevant.
   The second term of Eq.\ (\ref{dlag2}) is of ${\cal O}(p^4)$ and leads to a 
``modification'' of ${\cal L}_4$ \cite{Rudy_94,Scherer_95b},
\begin{equation}
\label{loffshell}
{\cal L}^{\mbox{\footnotesize off shell}}_4=
\beta_1 \mbox{Tr}({\cal O}_{EOM}^{(2)}
{\cal O}^{(2)\dagger}_{EOM})
+\beta_2\mbox{Tr}[(\chi U^\dagger-U\chi^\dagger){\cal O}_{EOM}^{(2)}],
\end{equation}
   where $\alpha_1=4\beta_1/F^2_0$ and 
$\alpha_2=-4(\beta_1+\beta_2)/F^2_0$ and $\beta_1$ and $\beta_2$ are 
now dimensionless.

   By a simple redefinition  of the field variables one generates an infinite 
set of ``new'' Lagrangians depending on two parameters $\beta_1$ and 
$\beta_2$.
   That all these Lagrangians describe the same physics will be illustrated
in the next section.
   In this sense we would argue that Eqs.\ (\ref{leff}) and (\ref{lagup})
represent the {\em same} theory in different representations.
   The concept of field transformations is very similar to choosing
appropriate coordinates in the description of a dynamical system.
   The value of physical observables should, of course, not depend
on the choice of coordinates.

\subsubsection{The Compton-scattering amplitude}
    The most general, irreducible, renormalized three-point Green's 
function [see Eq.\ (\ref{par})] at ${\cal O}(p^4)$
was derived in Ref.\ \cite{Rudy_94}. 
    For positively charged pions and for real photons ($q^2=0, q=p_f-p_i$) 
it has the simple form 
\begin{equation}
\label{emv}
\Gamma^{\mu,irr}_R(p_f,p_i)=(p_f+p_i)^{\mu}
\left(1+16\beta_1\frac{p_f^2+p_i^2-2M^2_{\pi}}{F^2_{\pi}}\right),
\end{equation}
and the corresponding renormalized propagator satisfying the Ward-Takahashi 
identity, Eq.\ (\ref{wti}), is given by 
\begin{equation}
\label{propagatoremv}
i\Delta_R(p)=\frac{i}{p^2-M^2_{\pi}
+\frac{16 \beta_1}{F^2_{\pi}}(p^2-M^2_{\pi})^2+i\epsilon}.
\end{equation}
   Clearly, the parameter $\beta_1$ is related to the deviation from a 
``pointlike'' vertex, once one of the pion legs is off shell. 
   Eqs.\ (\ref{emv}) and (\ref{propagatoremv}) have to be compared with 
the result of the usual representation of ChPT at ${\cal O}(p^4)$.
   In this case the vertex at $q^2=0$ is independent of $p_f^2$ and $p_i^2$,
$\Gamma^{\mu,irr}_R(p_f,p_i)=(p_f+p_i)^\mu$.
   Furthermore, the renormalized propagator is simply given by the free 
propagator.
 
   Let us now consider the process $\gamma(\epsilon,q)+\pi^+(p_i)
\to \gamma(\epsilon',q')+\pi^+(p_f)$.
   For $\beta_1\neq 0$ one expects a different contribution of the pole
terms, since the intermediate pion is not on its mass shell.   
   We subtract the ordinary calculation of the pole terms using free 
vertices from the corresponding calculation with off-shell
vertices and interpret the result as being due to off-shell effects.
   Similar methods have been the basis of investigating the influence of
off-shell form functions in various reactions involving the nucleon, such
as proton-proton bremsstrahlung \cite{Nyman_70,Kondratyuk_98} 
or electron-nucleus scattering \cite{Naus_87,Tiemeijer_90}.
   With the help of Eqs.\ (\ref{emv}) and (\ref{propagatoremv}) the change
in the pole amplitude can easily be calculated,\footnote{Of course, 
using Coulomb gauge 
$\epsilon^{\mu}=(0,\vec{\epsilon})$, $\epsilon'^{\mu}=(0,\vec{\epsilon}\,')$, 
and performing the calculation in the lab frame ($p_i^{\mu}=(M_{\pi},0)$),
the additional contribution vanishes, since $p_i\cdot\epsilon=
p_i\cdot\epsilon'=0$. 
   However, this is a gauge-dependent statement and thus not true for a 
general gauge.}
\begin{equation}
\label{dmp}
\Delta M_P=M_P(\beta_1\neq 0)-M_P(\beta_1=0)
=-ie^2 \frac{64 \beta_1}{F^2_{\pi}}
\left(
p_f\cdot\epsilon'\,p_i\cdot\epsilon+
p_f\cdot\epsilon\, p_i\cdot\epsilon'
\right).
\end{equation}
   However, Eq.\ (\ref{dmp}) cannot be used for a unique extraction of the 
form functions from experimental data since the very same term in the 
Lagrangian which contributes to the off-shell electromagnetic vertex also 
generates a two-photon contact interaction.
   This can be seen by inserting the appropriate covariant derivative into
Eq.\ (\ref{loffshell}) and by selecting those terms which contain two 
powers of the pion field as well as two powers of the electromagnetic field.
   From the first term of Eq.\ (\ref{loffshell}) one obtains the following
$\gamma\gamma\pi\pi$ interaction term 
\begin{eqnarray}
\label{loff1}
{\Delta\cal L}_{\gamma\gamma\pi\pi}&=&\frac{16\beta_1 e^2}{F^2_{\pi}}
\left\{-A^2[\pi^-(\Box+M^2_{\pi})\pi^+
+\pi^+(\Box+M^2_{\pi})\pi^-]\right.\nonumber\\
&&\left.+(\partial\cdot A+2A\cdot\partial)\pi^+
(\partial\cdot A+2A\cdot\partial)\pi^-\right\}.
\end{eqnarray}
   For real photons Eq. (\ref{loff1}) translates into a contact contribution 
of the form
\begin{equation}
\label{f1}
\Delta {\cal M}_{\gamma\gamma\pi\pi}=ie^2\frac{64\beta_1}{F^2_\pi}
(p_f\cdot\epsilon'\,p_i\cdot
\epsilon+p_f\cdot\epsilon\, p_i\cdot\epsilon'),
\end{equation}
which precisely cancels the contribution of Eq.\ (\ref{dmp}).
   At first sight the second term of Eq.\ (\ref{loffshell}) also seems to
generate a contribution to the Compton-scattering amplitude.
   However, after wave function renormalization this term drops out (see 
Ref.\ \cite{Scherer_95b} for details).
   We emphasize that all the cancellations happen only when one consistently 
calculates off-shell form functions, propagators and contact terms, and 
properly takes renormalization into account.
   Thus the Lagrangian of Eq.\ (\ref{lagup}) which represents an equivalent 
form to the standard Lagrangian of ChPT yields, as a consequence of the
equivalence theorem, the same Compton-scattering amplitude while, at the same 
time, it generates different off-shell form functions.
   Clearly, this illustrates why there is no unambiguous way of extracting 
the off-shell behavior of form functions from on-shell matrix elements.
   The ultimate reason is that the form functions of Eq.\ (\ref{par})
are not only model dependent but also representation dependent, {\em i.e.},
two representations of the same theory result in the same observables
but different form functions. 
 
\subsubsection{Off-shell effects versus contact interactions}
   In the present case there was a complete cancellation between 
off-shell effects and a contact contribution. 
   Even though this will not always necessarily be true, we would still
argue that as a matter of principle it is impossible to {\em uniquely} extract
off-shell effects from observables as there is a certain amount of
freedom to trade such effects for contact interactions.
   Let us discuss this claim within a somewhat different approach which 
does not make use of a calculation within a specific model or theory. 
   Such a discussion also serves to demonstrate that our interpretation is
independent of the fact that we made use of ChPT at ${\cal O}(p^4)$.  
   For that purpose we come back to the method of Gell-Mann and Goldberger 
in their derivation of the low-energy theorem for Compton scattering 
\cite{GellMann_54}, and split the most general invariant amplitude 
into two classes $A$ and $B$ (see Sec. III.B), where class $A$ consists of 
the most general pole terms and class $B$ contains
the rest.
   The original motivation in Ref.\ \cite{GellMann_54} for such a separation 
was to isolate those terms of ${\cal M}=-ie^2\epsilon_\mu\epsilon'^\ast_\nu
M^{\mu\nu}$ which have a singular behavior in 
the limit $q,q'\to 0$.
   As in Eq.\ (\ref{gammaa}) we write class $A$ in terms of the most general 
expressions for the irreducible, renormalized vertices and the renormalized 
propagator,
\begin{equation}
\label{ma}
M^{\mu\nu}_A=\Gamma^\nu(p_f,p_f+q')\Delta_R (p_i+q)
\Gamma^\mu (p_i+q,p_i)
+ (q\leftrightarrow -q',\mu\leftrightarrow\nu),
\end{equation}
where we made use of crossing symmetry.
   For sufficiently low energies class $B$ can be expanded in terms of the 
relevant kinematical variables [see Eq.\ (\ref{gammabansatz})].
   Furthermore, in class $A$ we expand the vertices and the renormalized pion 
propagator around their respective on-shell points, $p^2=M^2_\pi$.
   We obtain for the propagator
\begin{equation}
\label{exprop}
\Delta^{-1}_R(p)=p^2-M^2_\pi-\Sigma(p^2)=(p^2-M^2_\pi)
[1-\frac{p^2-M^2_\pi}{2}\Sigma''(M^2_\pi)+\cdots],
\end{equation}
where we made use of the standard normalization conditions 
$\Sigma(M^2_\pi)=\Sigma'(M^2_\pi)=0$.
   The expansion of, {\em e.g.}, the vertex describing the 
absorption of the initial real photon in the s channel reads 
\begin{eqnarray}
\label{expvert}
\Gamma^\mu(p_i+q,p_i)&=&(P^\mu+q'^\mu)F[0,M^2_\pi+(s-M^2_\pi),
M^2_\pi]\nonumber\\
&=&(P^\mu+q'^\mu)[1+(s-M^2_\pi)\partial_2F(0,M^2_\pi,M^2_\pi)+\cdots],
\end{eqnarray}
where $P=p_i+p_f$, and where $\partial_2$ refers to partial 
differentiation with respect to the second argument.
   Inserting the result of Eqs.\ (\ref{exprop}) and (\ref{expvert}) into 
Eq.\ (\ref{ma}) we obtain for the s-channel contribution to
$M^{\mu\nu}_A$
\begin{eqnarray}
\label{masexp}
M^{\mu\nu}_s&=&-ie^2 (P^\nu+q^\nu)[1+(s-M^2_\pi)
\partial_3 F(0,M^2_\pi,M^2_\pi) +\cdots]
\nonumber\\
&\times&\frac{1}{s-M^2_\pi}[1+\frac{s-M^2_\pi}{2}
\Sigma''(M^2_\pi)+\cdots]\nonumber\\
&\times&[P^\mu+q'^\mu)(1+(s-M^2_\pi)
\partial_2 F(0,M^2_\pi,M^2_\pi) +\cdots]\nonumber\\
&=&-ie^2\frac{(P^\nu+q^\nu)(P^\mu+q'^\mu)}{s-M^2_\pi}+
{\cal O}[(s-M^2_\pi)^0]\nonumber\\
&=&\mbox{``free'' s channel + analytical terms},
\end{eqnarray}
and an analogous term for the u channel.
   In Eq.\ (\ref{masexp}) ``free'' s channel refers to a calculation with
on-shell vertices.
   From Eq.\ (\ref{masexp}) we immediately see that off-shell effects 
resulting
from either the form functions or the renormalized propagator are of the same 
order as analytical contributions from class $B$.
   In the total amplitude off-shell contributions from class $A$ cannot 
uniquely be separated from class $B$ contributions.
   In the language of field transformations this means that contributions 
to $\cal M$ can be shifted between different diagrams leaving the total result
invariant. 
   In the language of Gell-Mann and Goldberger, by a change of representation,
contributions can be shifted from class $A$ to class $B$ within the {\em same}
theory.
   We can also express this differently; what appears to be an off-shell 
effect in one representation results, for example, from a contact interaction 
in another representation. 
    In this sense, off-shell effects are not only model dependent, 
{\em i.e.}, different models generate different off-shell form functions, 
but they are also representation dependent which means that even different 
representations of the {\em same} theory generate different off-shell form 
functions.
   This has to be contrasted with on-shell S-matrix elements which,
in general, will be different for different models (model dependent), 
but always the same for different representations of the same model
(representation independent). 
   For a further discussion in the context of bremsstrahlung the reader
is referred to Ref.\ \cite{Fearing_98b}.

\section{Virtual Compton scattering}
   In this section we will, finally, address an old topic 
\cite{Berg_61} which has recently
received considerable attention, namely, low-energy virtual Compton
scattering (VCS) as tested in, {\em e.g.}, the reactions 
$e^-p\to e^-p\gamma$ \cite{Brand_95,Audit_93,Brand_94,Audit_95,Shaw_97}
 and $\pi^- e^-\to\pi^-e^-\gamma$  \cite{Moinester_98a,Moinester_98b}.
   From a theoretical point of view, the objective of investigating
VCS is to map out all independent components of the Compton-scattering tensor,
Eq.\ (\ref{cst}), for arbitrary four-momenta of the photons.
   As we have seen, real Compton scattering is only sensitive to the
transverse components as well as restricted to the kinematics
$\omega=|\vec{q}|$ and $\omega'=|\vec{q}\,'|$.
   As in all studies with electromagnetic probes, the possibilities to
investigate the structure of the target increase substantially, if virtual
photons are used since (a) photon energy and momentum can be varied
independently and (b) longitudinal components of the transition current
are accessible.

\subsection{Kinematics and LET}

   When investigating $e^-p\to e^-p\gamma$ or $\pi^- e^-\to\pi^-e^-\gamma$,
the VCS tensor is only a building block of the invariant amplitude
describing the process.
   The total amplitude consists of a Bethe-Heitler (BH) piece,
where the real photon is emitted by the initial or final electrons,
and the VCS contribution (see Fig.\ \ref{fig:diagrams}),
\begin{equation}
{\cal M}={\cal M}_{\mbox{\footnotesize BH}}+{\cal M}_{\mbox{\footnotesize 
VCS}}.
\end{equation}
   It is straightforward to calculate the Bethe-Heitler contribution, since
it involves on-shell information of the target only, namely, its
electromagnetic form factors.
   In the following we will be concerned with the invariant amplitude for VCS.
    For the final photon the Lorentz condition $q'\cdot\epsilon'=0$ is
automatically satisfied, and we choose, in addition, the Coulomb
gauge $\epsilon'^\mu=(0,\vec{\epsilon}\,')$ implying
$\vec{q}\,'\cdot\vec{\epsilon}\,'=0$.
   Writing the invariant amplitude as 
$${\cal M}_{\mbox{\footnotesize VCS}}=-ie^2 \epsilon_\mu M^\mu,$$
where $\epsilon_\mu=e\bar{u}_e\gamma_\mu u_e/q^2$ is the polarization 
vector of the virtual photon, we can make use of current conservation, 
$q_\mu\epsilon^\mu=0$, $q_\mu M^\mu=0$, to express
$\epsilon_0$ and $M^0$ in terms of $\epsilon_z$ and $M_z$, respectively,
\begin{equation}
\label{mvcs}
{\cal M}_{\mbox{\footnotesize VCS}}
=ie^2\left(\vec{\epsilon}_T\cdot\vec{M}_T
+\frac{q^2}{\omega^2}\epsilon_z M_z\right).
\end{equation}
   Note that as $\omega\to 0$, both $\epsilon_z$ and $M_z$ tend to zero such
that ${\cal M}_{\mbox{\footnotesize VCS}}$ in Eq.\ (\ref{mvcs}) remains 
finite.

   After a reduction from Dirac spinors to Pauli spinors the transverse
and longitudinal parts of Eq.\ (\ref{mvcs}) may be expressed in terms
of 8 and 4 independent structures, respectively 
\cite{Guichon_95,Lvov_93,Scherer_96} (see Table \ref{numamp}):
\begin{eqnarray}
\label{mt}
\vec{\epsilon}_T\cdot\vec{M}_T&=&
\vec{\epsilon}\,'^\ast\cdot \vec{\epsilon}_T A_1
+i \vec{\sigma}\cdot(\vec{\epsilon}\,'^\ast\times\vec{\epsilon}_T) A_2
+(\hat{q}'\times\vec{\epsilon}\,'^\ast)\cdot(\hat{q}\times\vec{\epsilon}_T) A_3
+i \vec{\sigma}\cdot (\hat{q}'\times\vec{\epsilon}\,'^\ast)
\times(\hat{q}\times\vec{\epsilon}_T)A_4\nonumber\\
&&+i\hat{q}\cdot\vec{\epsilon}\,'^\ast
\vec{\sigma}\cdot(\hat{q}\times\vec{\epsilon}_T) A_5
+i\hat{q}'\cdot\vec{\epsilon}_T
\vec{\sigma}\cdot(\hat{q}'\times\vec{\epsilon}\,'^\ast) A_6
+i\hat{q}\cdot\vec{\epsilon}\,'^\ast
\vec{\sigma}\cdot(\hat{q}'\times\vec{\epsilon}_T) A_7\nonumber\\
&&+i\hat{q}'\cdot\vec{\epsilon}_T
\vec{\sigma}\cdot(\hat{q}\times\vec{\epsilon}\,'^\ast) A_8,\\
\label{ml}
\epsilon_z M_z&=&\epsilon_z \vec{\epsilon}\,'^\ast\cdot\hat{q} A_9
+i \epsilon_z\vec{\epsilon}\,'^\ast\cdot\hat{q}
\vec{\sigma}\cdot(\hat{q}'\times\hat{q})A_{10}
+i\epsilon_z\vec{\sigma}\cdot(\hat{q}\times\vec{\epsilon}\,'^\ast)A_{11}
+i\epsilon_z\vec{\sigma}\cdot(\hat{q}'\times\vec{\epsilon}\,'^\ast)A_{12},
\nonumber\\
\end{eqnarray}
   where the functions $A_i$ depend on three kinematical variables,
{\em e.g.}, $|\vec{q}|$, $\omega'=|\vec{q}\,'|$, and  
$z=\hat{q}\cdot\hat{q}\,'$.
      For the spin-zero case only one longitudinal and two transverse
structures are required.
   In Ref.\ \cite{Scherer_96} the method of Gell-Mann and Goldberger
was extended to VCS (see Sec. III.B) and model-independent predictions 
for the functions $A_i$ were obtained to second order in $q$ or $q'$.
   The results for the functions $A_i$ in the CM system expanded up
to ${\cal O}(2)$, {\em i.e.} $|\vec{q}\,'|^2$, 
$|\vec{q}\,'||\vec{q}|$ and $|\vec{q}|^2$,
are shown in Tables \ref{tablet} and \ref{tablel}.
   To this order, all $A_i$ can be expressed in terms of the proton mass
$M$, the anomalous magnetic moment $\kappa$, the electromagnetic Sachs
form factors $G_E$ and $G_M$, the mean square electric radius $r^2_E$, 
and the real-Compton-scattering electromagnetic polarizabilities
$\bar{\alpha}_E$  and $\bar{\beta}_M$.
   For $|\vec{q}|=\omega'$, the predictions of Table \ref{tablet} reduce to 
the well-known RCS result.   
   
   In Ref.\ \cite{Fearing_98} the low-energy behavior of
the VCS amplitude of $\pi^-(p_i)+\gamma^\ast(\epsilon,q)\to
\pi^-(p_f)+\gamma(\epsilon',q')$ was found to be of the form
\begin{equation}
\label{mvcspion}
{\cal M}_{\mbox{\footnotesize VCS}}=-2ie^2 F(q^2)\left[
\frac{p_f\cdot\epsilon'^\ast (2 p_i+q)\cdot\epsilon}{s-m_\pi^2}
+\frac{p_i\cdot\epsilon'^\ast (2 p_f-q)\cdot\epsilon}{u-m_\pi^2}
-\epsilon\cdot\epsilon'^\ast\right]+{\cal M}_R,
\end{equation}
where $F(q^2)$ is the electromagnetic form factor of the pion,
$s=(p_i+q)^2$, and $u=(p_i-q')^2$.
   The residual term ${\cal M}_R$ is separately gauge invariant and
at least of second order, {\em i.e.}, ${\cal O}(qq,qq',q'q')$.

\subsection{Beyond the LET: Generalized Polarizabilities}

   The framework for analyzing the model-dependent terms specific to VCS
at low energies has been developed by Guichon, Liu, and Thomas
\cite{Guichon_95}.
   The invariant amplitude ${\cal M}_{\mbox{\footnotesize VCS}}$ is split 
into a pole piece ${\cal M}_P$ and a residual part ${\cal M}_R$.  
   For the nucleon,  the s- and u-channel pole diagrams have been
calculated using the free Feynman propagator together with
electromagnetic vertices of the form
\begin{equation} 
\label{f1f2vertex}
\Gamma^\mu_{F_1F_2}(p_f,p_i)
=\gamma^\mu F_1(q^2)+i\frac{\sigma^{\mu\nu}q_\nu}{2M} F_2(q^2),
\,q=p_f-p_i, \end{equation}
where $F_1$ and $F_2$ are the Dirac and Pauli form factors, 
respectively.
   The corresponding amplitude ${\cal M}^{\gamma^\ast\gamma}_P$ contains all 
irregular terms as $q\to 0$ or $q'\to 0$ and is separately gauge invariant
\cite{Guichon_95,Scherer_96}.
   The second property is a special feature when working with these 
particular vertex operators and has been essential for the derivation in 
\cite{Guichon_95}. 
   Gauge invariance with this choice of operators is not as trivial as
it might appear, since the electromagnetic vertex, Eq.\ (\ref{f1f2vertex}), 
and the nucleon propagator do not satisfy the Ward-Takahashi identity 
(except at the real photon point,
$q^2 = 0)$:
\begin{equation}
\label{wtf1f2}
q_\mu \Gamma^\mu_{F_1,F_2}(p_f,p_i)=(p_f\hspace{-.9em}/\hspace{.3em}
-p_i\hspace{-.7em}/\hspace{.3em})F_1(q^2)\neq S_F^{-1}(p_f)
-S^{-1}_F(p_i).
\end{equation} 
   However, explicit calculation, including the use of the Dirac equation, 
shows that ${\cal M}_P$ of Ref.\ \cite{Guichon_95} is, in fact, identical with 
evaluating the pole terms using the vertex of Eq.\ (\ref{gammaeff}).
   We should mention that in Ref.\ \cite{Guichon_95} the phrase 
LET is used for the Born terms as opposed to the more restrictive sense of 
Sec.\ IV.A.

   For the pion, the situation is somewhat more complicated due to
the fact that even for real photons the s- and u-channel pole diagrams are 
not separately gauge invariant.
   A natural starting point is given by Eq.\ (\ref{mvcspion}).

   The generalized polarizabilities in VCS \cite{Guichon_95} result 
from an analysis of ${\cal M}_{R}^{\gamma^{\ast} \gamma}$ in terms of 
electromagnetic multipoles $H^{(\rho' L', \rho L)S}(\omega' , |\vec{q}|)$,  
where $\rho \, (\rho')$ denotes the type of the initial 
(final) photon ($\rho = 0$: charge, C; $\rho = 1$: magnetic, M;
$\rho = 2$: electric, E).  
   The initial (final) orbital angular momentum is denoted by 
$L \, (L')$, and $S$ distinguishes between non-spin-flip $(S = 0)$ and 
spin-flip $(S = 1)$ transitions.
   For the pion, only the case $S=0$ applies.
   ${\cal M}^{\gamma^\ast\gamma}_R$ is at least linear in the energy of the 
real photon.
   A restriction to the lowest-order, {\em i.e.} 
linear terms in $\omega'$ leads 
to only electric and magnetic dipole radiation in the final state.
   Parity and angular-momentum selection rules (see Table 
\ref{tab:mulan}) then allow for 3 scalar 
multipoles $(S = 0)$ and 7 vector multipoles $(S = 1)$. 
   The corresponding ten GPs, $P^{(01,01)0}$, ...,
${\hat{P}}^{(11,2)1}\,$,
are functions of $|\vec{q}|^2$, 
   where mixed-type polarizabilities, 
${\hat{P}}^{(\rho' L' , L)S} (|\vec{q}|^2)$, have been introduced which are 
neither purely electric nor purely Coulomb type (see Ref.\ 
\cite{Guichon_95} for details). 

   However, the treatment of Ref.\ \cite{Guichon_95} does not fully
exploit all the symmetries of the VCS tensor, resulting in redundant 
stuctures.  
   This observation was first made in Ref.\ \cite{Metz_96} in the
covariant framework of the linear sigma model.
    In fact, only six of the above ten GPs are independent, if 
charge-conjugation symmetry is imposed in combination with particle
crossing \cite{Drechsel_97,Drechsel_98b}.
   For example, for a charged pion, the constraint for the Compton
tensor reads \cite{Fearing_98,Drechsel_97}
\begin{equation}
\label{ccc}
M_{\pi^+}^{\mu\nu}(p_f,q';p_i,q)
\stackrel{C}{=}M_{\pi^-}^{\mu\nu}(p_f,q';p_i,q)
\stackrel{part.cross.}{=}M_{\pi^+}^{\mu\nu}(-p_i,q';-p_f,q),
\end{equation}
   generating one relation between originally three GPs 
\cite{Metz_96,Drechsel_97}:
\begin{equation}
\label{relpol}
0  =  \sqrt{\frac{3}{2}} P^{(01,01)0}(|\vec{q}|^2)
 + \sqrt{\frac{3}{8}} P^{(11,11)0}(|\vec{q}|^2)
 + \frac{3 |\vec{q}|^2}{2 \omega_0} \hat{P}^{(01,1)0}(|\vec{q}|^2),
\end{equation}
allowing one to eliminate the mixed-type polarizability
$\hat{P}^{(01,1)0}$.
   In Eq.\ (\ref{relpol}), $\omega_0=\left.\omega\right|_{\omega'=0}=
M-\sqrt{M^2+\vec{q}\,^2}$.
   The remaining two spin-independent polarizabilities have been
defined such as to be proportional to the RCS polarizabilities
at $|\vec{q}|=0$:
\begin{equation}
\bar{\alpha}_E=-\frac{e^2}{4\pi}\sqrt{\frac{3}{2}}
P^{(01,01)0}(0),\quad
\bar{\beta}_M=-\frac{e^2}{4\pi}\sqrt{\frac{3}{8}}
P^{(11,11)0}(0).
\end{equation}
   Note that a calculation in the framework of nonrelativistic quantum
mechanics is not particle-crossing symmetric and fails to satisfy the 
second equality in Eq.\ (\ref{ccc})
\cite{Pasquini_99} (see Sec.\ II.B).
   In general, we do not expect a nonrelativistic calculation 
to reproduce the constraints of \cite{Drechsel_97,Drechsel_98b}.

   Relations between the spin-dependent GPs at $|\vec{q}|=0$ and the four 
spin-dependent RCS polarizabilities $\gamma_i$ of Ref.\ 
\cite{Ragusa_93} were discussed in Ref.\ \cite{Drechsel_98c}: 
\begin{equation}
\label{rpgp}
\gamma_3 =  - \frac{e^2}{4 \pi}
\frac{3}{\sqrt{2}} P^{(01,12)1} (0),\quad 
\gamma_2 + \gamma_4 
=  - \frac{e^2}{4 \pi} \frac{3 \sqrt{3}}{2 \sqrt{2}} P^{(11,02)1}(0),
\end{equation}
{\em i.e.}, only two of the four $\gamma_i$ can be related to GPs 
at $|\vec{q}|=0$.

\subsection{Generalized Polarizabilities of the Nucleon}
   Predictions for the generalized polarizabilities of the nucleon have been
obtained in various approaches
\cite{Guichon_95,Metz_96,Liu_96,Vanderhaeghen_96,%
Hemmert_97a,Hemmert_97b,Metz_97,Kim_97,Pasquini_97,Korchin_98}.
   Here, we will concentrate on a discussion of the GPs obtained
within the heavy-baryon formulation 
of chiral perturbation theory (HBChPT) and the linear sigma model.
   Both calculations are based upon covariant approaches and thus 
satisfy the constraints found in \cite{Drechsel_97,Drechsel_98b}.

   The extension of meson chiral perturbation theory to the nucleon sector
starts from the most general effective chiral Lagrangian involving 
nucleons, pions, and external fields \cite{Gasser_88},
\begin{equation}
\label{leffnucl}
{\cal L}_{\mbox{\footnotesize eff}}
={\cal L}^{(1)}_{\pi N}+{\cal L}^{(2)}_{\pi N}+\cdots.
\end{equation}
   The lowest-order Lagrangian, 
\begin{equation}
\label{ln1}
{\cal L}_{\pi N}^{(1)} = \bar{\Psi}\left(i D\hspace{-.7em}/-m_0
+\frac{\stackrel{\circ}{g}_A}{2}u\hspace{-.6em}/\gamma_5\right)\Psi,
\end{equation}   
   contains two parameters in the chiral limit, namely, 
the axial-vector coupling constant $\stackrel{\circ}{g}_A$ and the nucleon 
mass $m_0$.
   The covariant derivatives $D_{\mu} \Psi$ includes, among other
terms, the coupling to the electromagnetic field, and $u_\mu$ contains in 
addition the derivative coupling of a pion.
   Since the nucleon mass does not vanish in the chiral limit, one has 
an additional large scale such that even in the chiral limit external 
four-momenta cannot be made arbitrarily small.
   In the framework of HBChPT \cite{Bernard_92,Jenkins_91} four-momenta
are written as $p = m_0 v +k$, $v^2=1$, $v^0\ge 1$, 
where often $v^\mu = (1,0,0,0)$ is used.
   By introducing so-called velocity-dependent fields 
\begin{displaymath}
{\cal N}_v=e^{+im_0v\cdot x}\frac{1}{2}(1+v\hspace{-.5em}/)\Psi,\quad
{\cal H}_v=e^{+im_0v\cdot x}\frac{1}{2}(1-v\hspace{-.5em}/)\Psi,
\end{displaymath}
such that $\Psi(x)=e^{-im_0 v \cdot x} ({\cal N}_v +{\cal H}_v)$,
   and by using the equation of motion for ${\cal H}_v$, one can
eliminate ${\cal H}_v$ to obtain a Lagrangian for ${\cal N}_v$
which, to lowest order in $1/m_0$, is given by
\begin{equation}
\label{lh1n}
\widehat{\cal L}^{(1)}_{\pi N}=\bar{\cal N}_v(iv\cdot D + g_A S\cdot u)
{\cal N}_v.
\end{equation}
   In Eq.\ (\ref{lh1n}) $S^{\mu} = i \gamma_5 \sigma^{\mu\nu} v_{\nu}$ 
refers to the spin matrix.
   This procedure can be generalized to higher orders in the chiral
expansion and is very smilar to the Foldy-Wouthuysen procedure
\cite{Foldy_50}.
   At each order in the momentum expansion one will have $1/m_0$ terms coming
from the expansion of the leading Lagrangian in combination with counter
terms resulting from the most general form allowed at that order.
  
\begin{figure}
\caption[test]{${\cal O}(p^3)$ prediction for the GPs of the proton
as a function of $\bar{q}^2=|\vec{q}|^2$ (from Ref.\ \cite{Hemmert_97b}).
   The dashed line represents the contribution from pion-nucleon
loops, the dotted line represents the $\pi^0$-exchange graph, and
the dash-dotted line represents the sum of both.
\label{gpschpt}}
\epsfig{file=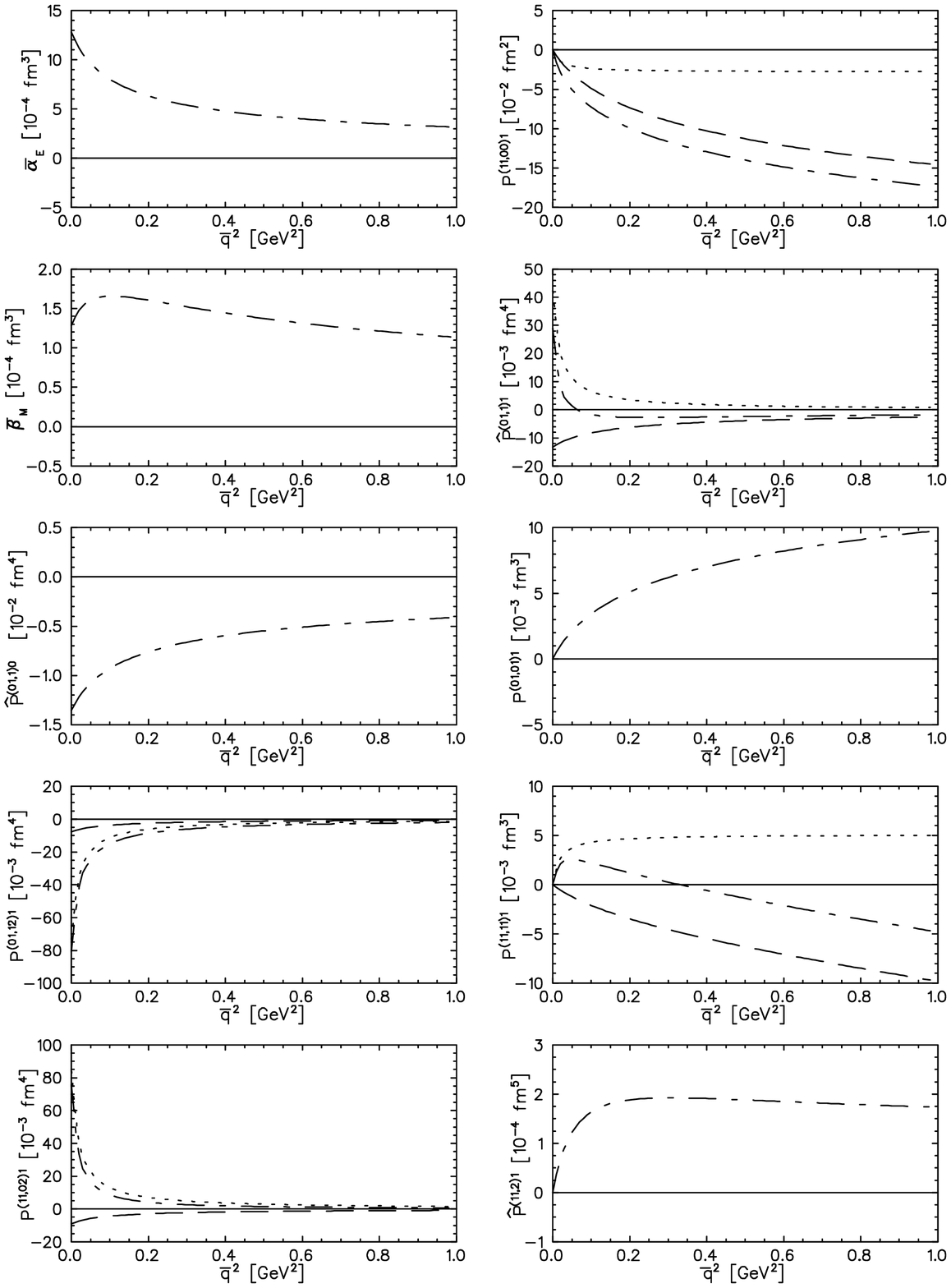,width=16cm}
\end{figure}

   In Refs.\ \cite{Hemmert_97a,Hemmert_97b} the VCS amplitude
was calculated using HBChPT to third order in the external momenta.
   At ${\cal O}(p^3)$, contributions to the GPs are generated by 
nine one-loop diagrams and the $\pi^0$-exchange $t$-channel pole graph (see
Fig.\ 2 of Ref.\ \cite{Hemmert_97a}). 
   For the loop diagrams only the leading-order Lagrangians of Eqs.\
(\ref{l2}) and (\ref{lh1n}) are needed.
   For the $\pi^0$-exchange diagram one requires, in addition, the 
$\pi^0\gamma\gamma^\ast$ vertex provided by the Wess-Zumino-Witten 
Lagrangian \cite{Wess_71,Witten_79},
\begin{equation}
\label{wzwpi0}
{\cal{L}}_{\gamma\gamma\pi^0}^{(WZW)} =  -\frac{e^2}{32\pi^2 F_\pi} \;
\epsilon^{\mu\nu\alpha\beta} F_{\mu\nu} F_{\alpha\beta} \pi^0 \,,
\end{equation}
   where $\epsilon_{0123}=1$ and $F_{\mu\nu}$ is the electromagnetic field
strength tensor.
   At ${\cal O}(p^3)$, the LET of VCS (see Tables \ref{tablet}
and \ref{tablel}) is reproduced by the 
tree-level diagrams obtained from Eq.\ (\ref{lh1n}) and the relevant 
part of the second- and third-order Lagrangian
\cite{Ecker_96},
\begin{eqnarray}
\label{lpin2}
\widehat{\cal L}^{(2)}_{\pi N}&=& - \frac{1}{2M} \bar N_v \left[
D \cdot D +\frac{e}{2} (\mu_S+\tau_3\mu_V)
\varepsilon_{\mu \nu \rho \sigma} F^{\mu\nu} v^{\rho} S^{\sigma}\right]
N_v,\\
\label{lpin3}
\widehat{\cal L}^{(3)}_{\pi N}&=&
\frac{ie\varepsilon_{\mu\nu\rho\sigma} F^{\mu\nu}}{8 M^2} \bar N_v
\left[\mu_S-\frac{1}{2}+\tau_3(\mu_V-\frac{1}{2})\right]
S^{\rho} D^{\sigma} N_v +h.c.\,.
\end{eqnarray}
   The numerical results for the ten GPs are shown in Fig.\ 
\ref{gpschpt} (recall that only six combinations are independent).
   As an example of the spin-independent sector, let us discuss the
generalized electric polarizability $\bar{\alpha}_E(|\vec{q}|^2)
=-\frac{e^2}{4\pi}\sqrt{\frac{3}{2}}
P^{(01,01)0}(|\vec{q}|^2)$,
\begin{equation}
\label{alphaq2}
\frac{\bar{\alpha}_E(\bar{q}^2)}{\bar{\alpha}_E}=
1-\frac{7}{50}\frac{|\vec{q}|^2}{m^2_\pi}
+\frac{81}{2800}\frac{|\vec{q}|^4}{m^4_\pi}
+O\left(\frac{|\vec{q}|^6}{m^6_\pi}\right),\,\,
\bar{\alpha}_E=\frac{5 e^2 g_A^2}{384\pi^2m_\pi F_\pi^2}
=12.8\times 10^{-4}\,\mbox{fm}^3.
\end{equation}
   As a function of $|\vec{q}|^2$, the prediction of ChPT decreases
considerably faster than within the constituent quark model \cite{Guichon_95}.
   At ${\cal O}(p^3)$, all results can be expressed in terms of the
pion mass $m_\pi$, the axial-vector coupling constant $g_A$, and
the pion-decay constant $F_\pi$.
   This property is true for all GPs at ${\cal O}(p^3)$.

   The $\pi^0$-exchange diagrams only contributes to the spin-dependent
polarizabilities.
   Let us consider as an example $P^{(11,11)1}$,
\begin{equation}
\label{p11111}
P^{(11,11)1}(|\vec{q}|^2)=-\frac{1}{288}\frac{g_A^2}{F_\pi^2}
\frac{1}{\pi^2 M}\left[\frac{|\vec{q}|^2}{m^2_\pi}-\frac{1}{10}
\frac{|\vec{q}|^4}{m^4_\pi}\right]
+\frac{1}{3M}\frac{g_A}{8\pi^2F^2_\pi}\tau_3
\left[\frac{|\vec{q}|^2}{m^2_\pi}-\frac{|\vec{q}|^4}{m^4_\pi}\right]
 +O\left(\frac{|\vec{q}|^6}{m^6_\pi}\right).
\end{equation}
   In general, the spin-dependent polarizabilities consist of an isoscalar
contribution due to pion loops [first term of Eq.\ (\ref{p11111})]
and an isovector contribution from
the $\pi^0$-exchange graph [second term of Eq.\ (\ref{p11111})].

\begin{figure}
\caption[test]{Some GPs of the proton. $Q_0^2=2m_N(\sqrt{|\vec{q}|^2
+m_N^2}-m_N)$.
Solid curve: linear sigma model 
\cite{Metz_96}. Dashed curve: Constituent
quark model \cite{Guichon_95}. Dotted curve: Effective
Lagrangian model \cite{Vanderhaeghen_96}.
Dashed-dotted curve: Chiral perturbation theory \cite{Hemmert_97a}.
\label{genpolcomp}}
\epsfig{file=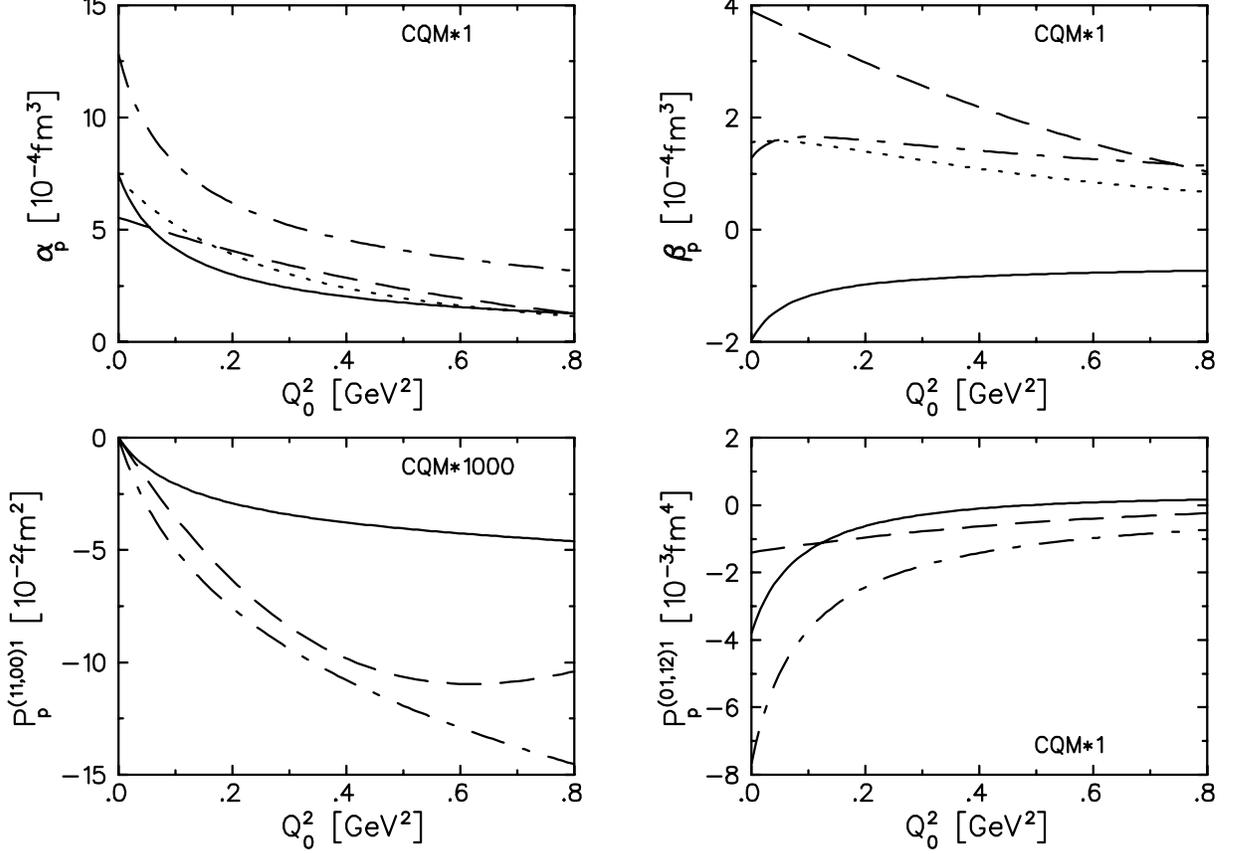,width=16cm}
\end{figure}

   The linear sigma model (LSM) \cite{GellMann_60}
represents a field-theoretical realization
of chiral $\mbox{SU(2)}_L\times\mbox{SU(2)}_R$ symmetry.
   The dynamical degrees of freedom are given by a nucleon doublet
$\Psi$, a pion triplet $\vec{\pi}$, and a singlet $\sigma$:
\begin{eqnarray}
\label{t:ss:lsm}
{\cal L}_S&=&i\bar{\Psi}\partial\hspace{-.5em}/\Psi 
+\frac{1}{2}\partial_\mu\sigma\partial^\mu\sigma+\frac{1}{2}
\partial_\mu\vec{\pi}\cdot\partial^\mu\vec{\pi}\nonumber\\
&&-g_{\pi N}\bar{\Psi}(\sigma+i\gamma_5\vec{\tau}\cdot\vec{\pi})\Psi
-\frac{\mu^2}{2}(\sigma^2+\vec{\pi}^2)
-\frac{\lambda}{4}(\sigma^2+\vec{\pi}^2)^2,\\ 
\label{t:ss:lsb}
{\cal L}_{s.b.}&=&-c\sigma,
\end{eqnarray}
   where ${\cal L}_{s.b.}$ is a perturbation which explicitly breaks
chiral symmetry.
   With an appropriate choice of parameters ($\mu^2 <0$, $\lambda >0$)
the model reveals spontaneous symmetry breaking, 
$<\!0|\sigma|0\!>=F_\pi=92.4$ MeV.
   The spectrum consists of massless pions, a massive sigma
and nucleons with masses satisfying the  
Goldberger-Treiman relation  $m_N=g_{\pi N} F_\pi$ with $g_A=1$.
   The symmetry breaking of Eq.\ (\ref{t:ss:lsb}) generates
the PCAC relation 
$$\partial^\mu A_\mu^a=F_\pi m^2_\pi \pi^a.$$
   The interaction with the electromagnetic field is introduced via
minimal substitution in Eq.\ (\ref{t:ss:lsm}).
   The generalized polarizabilities have been calculated in the
framework of a one-loop calculation \cite{Metz_96,Metz_97}.
   In Fig.\ \ref{genpolcomp}, some predictions of the LSM are shown
in comparison with other models.
   In Ref.\ \cite{Metz_97} for each generalized polarizability a chiral
expansion has been performed.
   As an example, consider $P^{(11,11)1}$ for the proton,
\begin{eqnarray}
\label{t:ss:p11111}
P^{(11,11)1}_p(Q^2_0)
&=&Q^2_0\frac{C}{2m_\pi}\left[\frac{6}{\mu}-\frac{1}{2\mu}+\frac{7\pi}{8}
+{\cal O}(\mu)\right]\nonumber\\
&&+(Q_0^2)^2 \frac{C}{5m_\pi^3}\left[-\frac{15}{\mu}+\frac{1}{8\mu}
-\frac{9\pi}{64}
+{\cal O}(\mu)\right]+{\cal O}[(Q_0^2)^3],
\end{eqnarray}
   where $C=g_{\pi N}^2/(72\pi^2 m_N^4)$, $\mu=m_\pi/m_N$ und $Q_0^2=2 m_N
(\sqrt{|\vec{q}|^2+m_N^2}-m_N)$.
   The leading-order term of the chiral expansion agress with the 
corresponding result of HBChPT at ${\cal O}(p^3)$.

\subsection{Generalized Polarizabilities of Pions}
   Of course, the concept of (generalized) polarizabilities can also
be applied to the pion.
   From the experimental point of view, an extraction of polarizabilities
is more complicated since there is no pion target.
   For that purpose, one has to consider reactions which contain 
the Compton-scattering amplitude as a building block, such as, {\em e.g.},
the Primakoff effect in high-energy pion-nucleon bremsstrahlung,
$\pi^-A\to A\pi^-\gamma$ \cite{Antipov_83,Antipov_85},
radiative pion photoproduction off the nucleon,
$\gamma p\to n\gamma \pi^+$ \cite{Aibergenov_86},
and pion pair production in $e^+e^-$ scattering, $e^+e^-\to
e^+e^-\pi^+\pi^-$ \cite{MARKII_90,CELLO_92}.
   The current empirical information on the RCS polarizabilities
is summarized in Table \ref{pionpolemp}.
   New experiments are presently beeing performed \cite{Ahrens_95}
or in preparation \cite{Gorringe_98}.

   From a theoretical point of view, a precise determination of the
electromagnetic polarizabilities is of great importance as a test of
ChPT. 
   At the one-loop level, ${\cal O}(p^4)$, of chiral perturbation 
theory the electromagnetic 
polarizabilities of the charged pion are entirely determined by
an ${\cal O}(p^4)$ counter term \cite{Holstein_90}, 
\begin{equation}
\label{t:ss:alpha}
\bar{\alpha}_E=-\bar{\beta}_M
=\frac{e^2}{4\pi} \frac{2}{m_\pi F^2_\pi}(2l^r_5-l^r_6)
=(2.68\pm0.42)\times 10^{-4}\,\mbox{fm}^3,
\end{equation}
   where the linear combination $2l^r_5-l^r_6$ is predicted through
the decay $\pi^+\to e^+\nu_e\gamma$.
   Corrections to this result at ${\cal O}(p^6)$ were shown
to be reasonably small, namely 12\% and 24\% of the ${\cal O}(p^4)$ values
for $\bar{\alpha}_E$ and $\bar{\beta}_M$, respectively \cite{Buergi_96}.
   In particular, the degeneracy $|\bar{\alpha}_E|=|\bar{\beta}_M|$
of Eq.\ (\ref{t:ss:alpha}) is lifted at ${\cal O}(p^6)$.
   
\begin{figure}
\caption{Bethe-Heitler diagrams (a) and (b). VCS diagram (c).
The four-momenta of the virtual photons in the BH diagrams and the
VCS diagram differ from each other.
\label{fig:diagrams}}\begin{center}
\epsfig{file=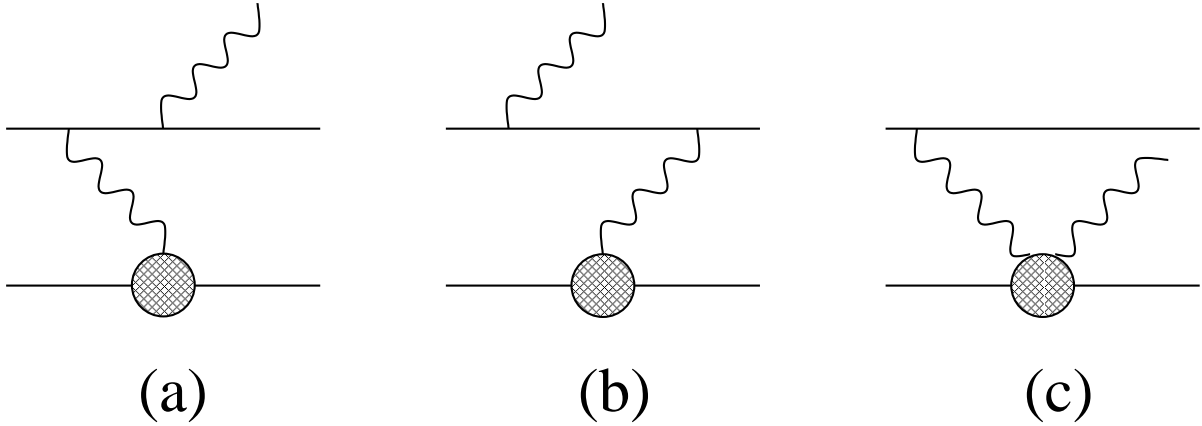,width=16cm}
\end{center}
\end{figure}

   Presently, the pion VCS reaction is under investigation as part of the
Fermi\-lab E781 SELEX experiment, where a 600 GeV pion beam interacts with 
atomic electrons in nuclear targets \cite{Moinester_98b}.    
   In principle, the different behavior under the substitution
$\pi^-\to\pi^+$ of ${\cal M}_{\mbox{\footnotesize BH}}$ and 
${\cal M}_{\mbox{\footnotesize VCS}}$ (see Fig.\ \ref{fig:diagrams}), 
\begin{equation}
{\cal M}_{\mbox{\footnotesize BH}}(\pi^-)= 
-{\cal M}_{\mbox{\footnotesize BH}}(\pi^+),\quad
{\cal M}_{\mbox{\footnotesize VCS}}(\pi^-)
= {\cal M}_{\mbox{\footnotesize VCS}}(\pi^+),
\end{equation}
   may be of use in identifying the contributions due to internal
structure by comparing the reactions involving a $\pi^-$ and a $\pi^+$ beam
for the same kinematics:\footnote{This argument works for any particle
which is not its own antiparticle such as the $K^+$ or $K^0$. 
Of course, one could also employ the substitution $e^-\to e^+$.} 
\begin{equation}
d\sigma(\pi^+)-d\sigma(\pi^-)\sim 4 Re \left(
{\cal M}_{\mbox{\footnotesize BH}}(\pi^+)
{\cal M}^\ast_{\mbox{\footnotesize VCS}}(\pi^+)\right).
\end{equation}

\begin{figure}
\caption[test]{Generalized polarizability $\bar{\alpha}_E(|\vec{q}|^2)$ 
of the charged pion in ChPT \cite{Unkmeir_99} .
Note that $\bar{\alpha}_E(|\vec{q}|^2)
=-\bar{\beta}_M(|\vec{q}|^2)$ at ${\cal O}(p^4)$.
\label{fig:alphapion}}
\epsfig{file=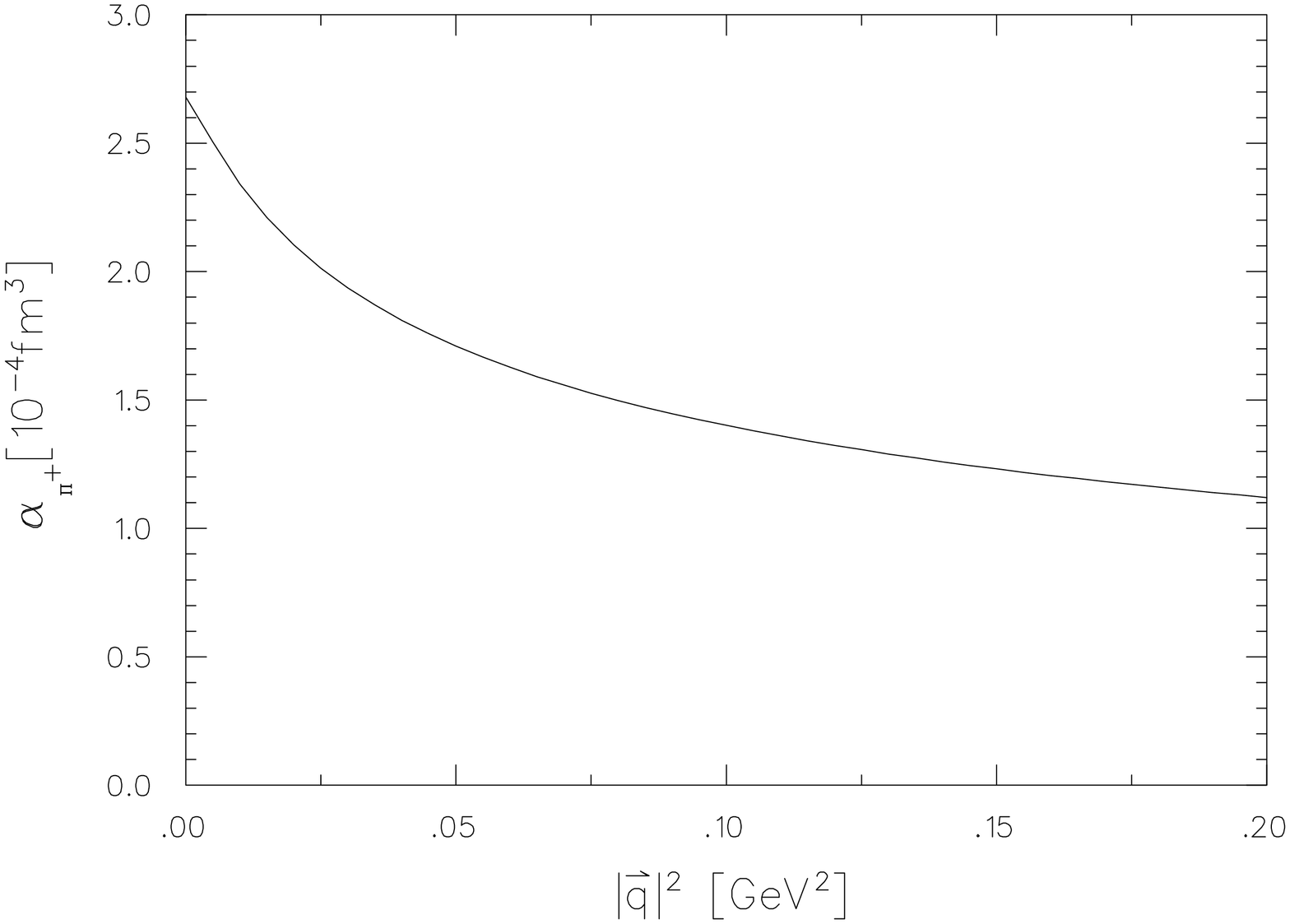,width=8cm,angle=90}
\end{figure}

   The invariant amplitude for VCS at ${\cal O}(p^4)$ in the framework of 
chiral perturbation theory has been calculated in Ref. \cite{Unkmeir_99}.
   The result was found to be in complete agreement with Eq.\ 
(\ref{mvcspion}).
   Using the procedure developed in Ref.\ \cite{Drechsel_97},
the generalized polarizabilities of the charged pion in the
convention of Ref.\ \cite{Guichon_95} were extracted,
\begin{eqnarray}
\label{t:ss:genalpha}
\bar{\alpha}_E(|\vec{q}|^2)&=&-\bar{\beta}_M(|\vec{q}|^2)\nonumber\\
&=&
\frac{e^2}{8\pi m_\pi}\sqrt{\frac{m_\pi}{E_\pi}}
\left[\frac{4(2l^r_5-l^r_6)}{F^2_\pi}-2 \frac{m_\pi-E_\pi}{m_\pi}
\frac{1}{(4\pi F)^2} {J^{(0)}}'\left(2\frac{m_\pi-E_\pi}{m_\pi}\right)\right],
\nonumber\\
\end{eqnarray}
where $E_\pi=\sqrt{m_\pi^2+|\vec{q}|^2}$ and 
$${J^{(0)}}'(x)=\frac{1}{x}\left[1-\frac{2}{x\sigma(x)}\ln\left(
\frac{\sigma(x)-1}{\sigma(x)+1}\right)\right],
\quad \sigma(x)=\sqrt{1-\frac{4}{x}},
\quad x\le 0.
$$
   The $|\vec{q}|^2$ dependence does not contain any ${\cal O}(p^4)$ 
parameter, {\em i.e.}, it is entirely given in terms of the pion mass 
and the pion decay constant $F_\pi=92.4$ MeV.
   The numerical prediction is shown in Fig.\ \ref{fig:alphapion}.

\acknowledgements
   I would like to thank the organizers, in particular, Petr Bydzovsky
for a very efficient organization and a pleasant atmosphere at
the summer school. 
   The author would like to thank D.\ Drechsel, H.W.\ Fearing, T.R.\ Hemmert,
B.R.\ Holstein, G.\ Kn\"{o}chlein, J.H.\ Koch, A.Yu.\ Korchin, A.I.\ L'vov,
A.\ Metz, B. Pasquini, and C.\ Unkmeir for a pleasant and fruitful 
collaboration on various topics related to virtual Compton scattering. 
   It is pleasure to thank J.M.\ Friedrich, N.\ d'Hose, M.A.\ Moinester,
and A.\ Ocherashvili for useful discussions on experimental issues
in VCS.
   Finally, many thanks are due to B.\ Pasquini for carefully reading the
manuscript.


\begin{table}
\caption[test]{\label{tcs} Thomson cross section $\sigma_T$ for
the electron, charged pion, and the proton.}
\begin{tabular}{|c|c|}
\hline
particle&$\sigma_T$\\
\hline
electron & 0.665 barn\\
\hline
pion & 8.84 $\mu$barn\\
\hline
proton & 197 nbarn\\
\hline
\end{tabular}
\end{table}

\begin{table}
\caption[test]{\label{rcspoln} 
Some theoretical predictions for the electromagnetic polarizabilities
of the nucleon in units of $10^{-4}\,\mbox{fm}^3$.}
\begin{tabular}{|c|c|c|c|c|c|}
\hline
Ref.&
description&
$\bar{\alpha}_E$&
$\bar{\beta}_M$&
$\bar{\alpha}_E$&
$\bar{\beta}_M$\\
&&proton&proton&neutron&neutron\\
\hline
\cite{Hecking_81}&MIT Bag &7.1&2.6&4.7&3.4\\ \hline
\cite{Nyman_84}&Skyrme model&&2&&2\\ \hline
\cite{Schaefer_84}&MIT Bag &10.8&2.3&10.8&1.5\\ \hline
\cite{Weiner_85}&Chiral Quark Model&7-9&$\leq 2$&7-9&$\leq 2$\\ \hline
\cite{Chemtob_87}&Skyrme&8.3&8.5&8.3&8.5\\ 
&&25.2&1.7&25.2&1.7\\ \hline
\cite{Scoccola_89}&Chiral Soliton&13.4&-1.1&13.4&-1.1\\ \hline
\cite{Bernard_92}&HBChPT ${\cal O}(p^3)$&12.8&1.3&12.8&1.3\\
\hline 
\cite{Li_93}&NRQM&7.25&12&7.25&12\\
\hline
\cite{Hemmert_97}&$\epsilon^3$ ChPT&17.1&9.2&17.1&9.2\\ \hline
\end{tabular}
\end{table}

\begin{table}
\caption[test]{\label{rcspolpemp} 
   Empirical numbers for the electromagnetic polarizabilities
of the proton in units of $10^{-4}\,\mbox{fm}^3$.
   The polarizabilities have been determined using the Baldin
sum rule, Eq.\ (\ref{baldinsumrule}), with
$\bar{\alpha}_p+\bar{\beta}_p=14.2\pm 0.5$.
   Due to this constraint the errors of $\bar{\alpha}_p$ and
$\bar{\beta}_p$ are anticorrelated.
}
\begin{tabular}{|c|c|c|c|}
\hline
Ref.&
description&
$\bar{\alpha}_E^p$&$\bar{\beta}_M^p$\\
\hline
\cite{Federspiel_91}&$\gamma p\to \gamma p$&
$10.9\pm 2.2\pm 1.3$&$3.3\pm 2.2\pm 1.3$\\
\cite{Zieger_92}&$\gamma p\to \gamma p$&
$10.62^{+1.25+1.07}_{-1.19-1.03}$&
$3.58^{+1.19+1.03}_{-1.25-1.07}$\\
\cite{Hallin_93}&$\gamma p\to \gamma p$&
$9.8\pm 0.4\pm 1.1$&$4.4\pm 0.4\pm 1.1$\\
\cite{MacGibbon_95}&$\gamma p\to\gamma p$&
$12.5\pm 0.6\pm 0.9$&$1.7\pm 0.6\pm 0.9$\\
\cite{MacGibbon_95}&global average&
$12.1\pm 0.8\pm 0.5$&$2.1\pm 0.8\pm 0.5$\\
\hline
\end{tabular}
\end{table}

\begin{table}
\caption[test]{\label{rcspolnemp} 
Empirical numbers for the electric polarizability
of the neutron in units of $10^{-4}\,\mbox{fm}^3$.}
\begin{tabular}{|c|c|c|}
\hline
Ref.&
description&$\bar{\alpha}_E^n$\\
\hline
\cite{Schmiedmayer_88}&low-energy $n$Pb and $n$C scattering
&$12\pm 10$\\
\cite{Koester_88}&low-energy $n$Pb and $n$Bi scattering
&$8\pm 10$\\
\cite{Rose_90a}&
quasi-free Compton scattering: 
$\gamma d\to \gamma'np$&
$11.7^{+4.3}_{-11.7}$\\
\cite{Rose_90b}&
quasi-free Compton scattering: 
$\gamma d\to \gamma'np$&
$10.7^{+3.3}_{-10.7}$\\
\cite{Schmiedmayer_91}&
low-energy $n$Pb scattering&
$12.0\pm 1.5\pm 2.0$\\
\cite{Koester_95}&low-energy $n$Pb scattering&$0 \pm 5$\\
\cite{PDG_98}&PDG average&$9.8^{+1.9}_{-2.3}$\\
\hline
\end{tabular}
\end{table}

\begin{table}
\caption[test]{\label{tablet}
   Transverse functions $A_i$ of Eq.\ (\ref{mt}) in the CM frame.
   The functions are expanded in terms of $|\vec{q}\,'|$ and $|\vec{q}|$
of the final real and initial virtual photon, respectively.
$N_i=\sqrt{\frac{E_i+M}{2M}}$ is the normalization factor of the initial
spinor, where $E_i=\sqrt{M^2+|\vec{q}|^2}$.
   $G_E(q^2)=F_1(q^2)+\frac{q^2}{4M^2}F_2(q^2)$ and $G_M(q^2)=F_1(q^2)
+F_2(q^2)$ are the electric and magnetic Sachs form factors,
respectively.
   $r^2_E=6 G_E'(0)=(0.74\pm 0.02)\, \mbox{fm}^2$ is the electric mean square
radius \cite{Simon_80} and $\kappa=1.79$ the anomalous magnetic moment of the
proton.
   $Q^\mu$ is defined as $q^\mu|_{|\vec{q}\,'|=0}=(M-E_i,\vec{q})$,
$Q^2=-2M(E_i-M)$, and $z=\hat{q}\,'\cdot\hat{q}$.
$\bar{\alpha}_E$ and $\bar{\beta}_M$ are the electric and magnetic Compton
polarizabilities of the proton, respectively.}
\begin{tabular}{c|c}
$A_1$&
$-\frac{1}{M}+\frac{z}{M^2}|\vec{q}|
-\left(\frac{1}{8M^3}+\frac{r^2_E}{6M}-\frac{\kappa}{4M^3}
-\frac{4\pi\bar{\alpha}_E}{e^2}\right)
|\vec{q}\,'|^2
+\left(\frac{1}{8M^3}+\frac{r^2_E}{6M}-\frac{z^2}{M^3}+
\frac{(1+\kappa)\kappa}{4M^3}\right)|\vec{q}|^2$\\
\hline
$A_2$&
$\frac{1+2\kappa}{2M^2}|\vec{q}\,'|
-\frac{\kappa^2}{4M^3}|\vec{q}\,'|^2
+\frac{z\kappa}{2M^3}|\vec{q}\,'||\vec{q}|
-\frac{(1+\kappa)^2}{4M^3}|\vec{q}|^2$\\
\hline
$A_3$&
$-\frac{1}{M^2}|\vec{q}|
+\left(\frac{1}{4M^3}+\frac{4\pi \bar{\beta}_M}{e^2}
\right)|\vec{q}\,'||\vec{q}|
+\frac{(3-2\kappa-\kappa^2)z}{4M^3}|\vec{q}|^2$\\
\hline
$A_4$&
$-\frac{(1+\kappa)^2}{2M^2}|\vec{q}|
-\frac{(2+\kappa)\kappa}{4M^3}|\vec{q}\,'||\vec{q}|
+\frac{(1+\kappa)^2 z}{4M^3}|\vec{q}|^2$\\
\hline
$A_5$&
$-\frac{N_i G_M(Q^2)}{(E_i+z|\vec{q}|)(E_i+M)}
\frac{|\vec{q}|^2}{|\vec{q}\,'|}
+\frac{(1+\kappa)\kappa}{4M^3}|\vec{q}|^2$\\
\hline
$A_6$&
$\frac{1+\kappa}{2M^2}|\vec{q}\,'|
-\frac{(1+\kappa)\kappa}{4M^3}|\vec{q}\,'|^2
-\frac{(1+\kappa)z}{2M^3}|\vec{q}\,'||\vec{q}|$\\
\hline
$A_7$&
$-\frac{1+3\kappa}{4M^3}|\vec{q}\,'||\vec{q}|$\\
\hline
$A_8$&
$\frac{1+3\kappa}{4M^3}|\vec{q}\,'||\vec{q}|$
\end{tabular}
\end{table}

\begin{table}
\caption[test]{\label{tablel} Longitudinal functions $A_i$ of Eq.\ (\ref{ml})
in the CM frame. See caption of Table \ref{tablet}.}
\begin{tabular}{c|c}
$A_9$&
$\frac{N_i G_E(Q^2)}{(E_i+z|\vec{q}|)(E_i+M)}\frac{|\vec{q}|^2}{|\vec{q}\,'|}
-\frac{1}{M}
+\frac{z}{M^2}|\vec{q}|
-\left(\frac{1}{8M^3}+\frac{r^2_E}{6M}-\frac{\kappa}{4M^3}
-\frac{4\pi\bar{\alpha}_E}{e^2}\right)
|\vec{q}\,'|^2
+\left(\frac{1}{8M^3}+\frac{r^2_E}{6M}-\frac{z^2}{M^3}\right)
|\vec{q}|^2$\\
\hline
$A_{10}$&
$-\frac{1+3\kappa}{4M^3}|\vec{q}\,'||\vec{q}|$\\
\hline
$A_{11}$&
$-\frac{1+2\kappa}{2M^2}|\vec{q}\,'|
+\frac{\kappa^2}{4M^3}|\vec{q}\,'|^2
+\frac{(1+\kappa)z}{4M^3}|\vec{q}\,'||\vec{q}|
+\frac{1+2\kappa}{4M^3}|\vec{q}|^2$\\
\hline
$A_{12}$&
$\frac{(1+\kappa)z}{2M^2}|\vec{q}\,'|
-\frac{(1+\kappa)\kappa z}{4M^3}|\vec{q}\,'|^2
-\frac{(1+\kappa)(2z^2-1)}{4M^3}|\vec{q}\,'||\vec{q}|
-\frac{(1+\kappa)z}{4M^3}|\vec{q}|^2$
\end{tabular}
\end{table}

\begin{table}
\caption[test]{\label{numamp} 
Number of independent amplitudes for the description of 
the general VCS tensor.}
\begin{tabular}{|l|c|c|c|}
\hline
nucleon& RCS ($\gamma\gamma$)&VCS ($\gamma^\ast\gamma$)
&VCS ($\gamma^\ast\gamma^\ast$)\\
\hline
total number of amplitudes&6&12&18\\
\hline
spin independent ($=$ pion)&2&3&5\\
\hline
spin dependent&4&9&13\\
\hline
\end{tabular}
\end{table}

\begin{table}
\caption[test]{Multipolarities of initial and final states.}
\label{tab:mulan}
\begin{tabular}
{|c|c|c|}
\hline
$J^P$&final real photon&initial virtual photon\\
\hline
$\frac{1}{2}^-$&E1&C1,E1\\
\hline
$\frac{3}{2}^-$&E1&C1,E1,M2\\
\hline
\hline
$\frac{1}{2}^+$&M1&C0,M1\\
\hline
$\frac{3}{2}^+$&M1&C2,E2,M1\\
\hline
\end{tabular}
\end{table}

\begin{table}
\caption[test]{Empirical numbers for the electromagnetic polarizabilities
of the pion in units of $10^{-4}\,\mbox{fm}^3$.
\label{pionpolemp}}
\begin{tabular}{|c|c|c|c|}
\hline
reaction & $\bar{\alpha}_E$ &
$\bar{\beta}_M$ &
$\bar{\alpha}_E+\bar{\beta}_M$
\\ \hline
$\pi^-A\rightarrow A\pi^-\gamma$ &$6.8\pm 1.4 \cite{Antipov_83}$ &
$-7.1 \pm 2.8_{\mbox{\footnotesize stat.}}
\pm 1.8_{\mbox{\footnotesize syst.}}$ \cite{Antipov_85}&
$1.4\pm 3.1_{\mbox{\footnotesize stat.}}\pm 2.5_{\mbox{\footnotesize syst.}}$
\cite{Antipov_85}
\\\hline
$\gamma p \rightarrow \gamma \pi^+ n$ &$20\pm 12$ 
\cite{Aibergenov_86}
&-&- 
\\ \hline\end{tabular}
\end{table}

\end{document}